\documentclass[11pt]{article}

\usepackage{graphicx,pstricks}
\usepackage{moreverb}
\usepackage{subfigure}
\usepackage{amsmath}
\usepackage{mathrsfs}

\usepackage{txfonts}
\usepackage{palatino}

\usepackage{xfrac}
\usepackage{mdwlist}
\usepackage{morefloats}
\usepackage{cite}
\usepackage{multirow}
\usepackage{authblk}

\DeclareMathAlphabet{\mathsfsl}{OT1}{cmss}{m}{sl}

\DeclareMathOperator{\trace}{tr}

\newcommand{\fepx}{{\bfseries{\slshape{FEpX}}}}

\newcommand{\vctr}[1]{\boldsymbol{#1}}

\newcommand{\tnsr}[1]{\mathsfsl{#1 }}

\newcommand{\gtnsr}[1]{\mathsfsl{#1}}

\newcommand{\gradop}{{\rm grad }}
\newcommand{\divop}{{\rm div}}

\newcommand{\transp}{{\rm T}}
\newcommand{\invrs}{{ -\hspace{-.1em}1}}
\newcommand{\deeteei }{ \frac{1}{\Delta t}  }   
\newcommand{\betaoverkappa}{ \frac{\beta}{\kappa \Delta t}  }

\newcommand{\cauchy}{{\boldsymbol{\sigma}}}
\newcommand{\dcauchy}{{\boldsymbol{\sigma}}^\prime}

\newcommand{\pxspinhat}{\hat{\tnsr{w}}^{p}}

\newcommand{\rss}{\tau^\alpha}

\newcommand{\sumss}{\sum_{\alpha}}
\newcommand{\matlatepsdot}{ \left\{ \dot{ \mathsf{e}^e} \right\} }   
   
\newcommand{\matlateps }{ \left\{\mathsf{e}^e\right\}  }   
\newcommand{\matlatepsold }{ \left\{\mathsf{e}_0^e\right\}  }  
\newcommand{\matdlateps }{ \left\{ {\mathsf{e}^e}^\prime \right\}  }    
\newcommand{\matdlatepsold }{ \left\{ {\mathsf{e}_0^e}^\prime \right\}  }  
\newcommand{\matdefrate }{\Big\{ \mathsf{d} \Big\}   }     
  
\newcommand{\matddefrate }{\Big\{ \mathsf{d}^\prime \Big\}   }
\newcommand{\matlatdefrate }{\Big\{ {\hat{\mathsf{d}}^p} \Big\}   }
\newcommand{\matpxspinhat }{\Big[ {\hat{\mathsf{w}}^p} \Big]   }
  
\newcommand{\matdkirch }{  \left\{ \tau^\prime \right\}  }   
  
\newcommand{\matdcauchy }{  \left\{ \sigma^\prime \right\}  } 
\newcommand{\matxdelasticity}{\Big[ \,\mathsf{c}^\prime \,\Big]}  
  
\newcommand{\matxplasticity}{\Big[ \,\mathsf{m} \,\Big]}  
\newcommand{\matxep}{\Big[ \,\mathsf{s} \,\Big]}

\newcommand{\matsymschmid }{ \Big\{\mathsf{p}^\alpha\Big\}  }  
\newcommand{\mathhh}{\Big\{ \mathsf{h} \Big\}   }   
   
\newcommand{\matbodyforce}{\Big\{ \iota \Big\}   }

\newcommand{\matvel }{ \Big\{ v \Big\}  } 
\newcommand{\matcapX}{\Big[ \,\mathsf{X} \,\Big]}    
\newcommand{\matcapB}{\Big[ \,\mathsf{B} \,\Big]}
\newcommand{\matcapN}{\Big[ \,\mathsf{N}(\xi, \eta, \zeta) \,\Big]}

\newcommand{\matresiduale}{\Big\{ R^{\it ele}_u \Big\}}
\newcommand{\matvelnp}{\Big\{ \mathsf{V} \Big\}}

\newcommand{\matstiffd}{\Big[ \,\mathsf{k}^{\it ele}_d \,\Big]} 
\newcommand{\matstiffv}{\Big[ \,\mathsf{k}^{\it ele}_v \,\Big]}

\newcommand{\matdelta}{\Big\{ \mathsf{\delta} \Big\}}

\newcommand{\dee}{{\mathrm{d}}}

\begin{document}
\title{Sensitivity of Crystal Stress Distributions\\ to  the Definition of \\ Virtual Two-Phase Microstructures}
\author[1]{Andrew C. Poshadel}
\author[2]{Michael Gharghouri}
\author[1]{Paul R. Dawson}
\affil[1]{Sibley School of Mechanical and Aerospace Engineering, Cornell University, Ithaca, New York, USA}
\affil[2] {Canadian Nuclear Laboratories, Chalk River, Ontario, Canada}
\date{}

\maketitle

\begin{abstract}
A systematic study of the sensitivities of simulation input  on the computed stress distributions in two-phase microstructures is presented.  The study supports an investigation of the initiation and propagation of yielding in duplex stainless steel~\cite{pos_daw_twophase}.  Considered in the study are the identification of constitutive model parameters for the single-crystal elastic and plastic behaviors and the importance of including dominant phase and grain morphologies in the instantiation of virtual samples.  Behaviors computed using a finite element formulation are evaluated against experimental data for the macroscopic stress-strain behavior and against lattice strain data measured by neutron diffraction under {\it in situ} loading.  
\end{abstract}
\pagebreak[4]

\section{Introduction}
\label{sec:introduction}
 Duplex steels, like many other structural alloys, are  complex materials that have been developed with performance in mind.  Alloying and processing have been optimized to achieve excellent mechanical properties by emphasizing favorable attributes of the microstructure.   However, quantitative links between microstructural attributes 
 and derivative mechanical properties often are difficult to establish.   
 Consequently,  efforts to model the mechanical behaviors are stymied by uncertainties in specifying the 
 microstructural state and crystal properties of virtual samples.  
 For simulations at the crystal scale, uncertainty enters in the constitutive parameters and in the instantiation of virtual samples. 
  In this paper we  address both of these sources of uncertainty for the duplex steel, LDX2101. 
  The study reported here is a complement to a paper focused on the initiation and propagation of yielding in LDX-2101~\cite{pos_daw_twophase}. 
  In that paper, the use of a multiaxial strength-to-stiffness parameter in predicting yielding is investigated.  
  Biaxial loading conditions are examined via comparisons between simulations and coordinated, {\it in situ} neutron diffraction measurements.   
   The present paper documents  a comprehensive  study  to quantify the sensitivities of the predicted responses
to constitutive parameter identification and sample instantiation and thereby to offer confidence in the predictive
capabilities demonstrated in \cite{pos_daw_twophase}.   

The investigation is presented as two semi-independent studies.  
First, we consider the constitutive model parameters.  As the modeling is conducted at the scale of crystal aggregates in which individual crystals are spatially resolved with finite elements, the parameters are those associated with single crystals of the two phases (austenite and ferrite).  For the elastic response, we require the moduli appearing in the anisotropic Hooke's law.  For the plastic response, we consider restricted slip as the dominant mode of plastic deformation, and thus need the parameters associated with the kinetics of slip at fixed state (also referred to as the flow law) and the parameters appearing in the equation for strain hardening.  Here, we employ the same form of a modified Voce expression for both phases.   Besides the evolution equation parameters, we also examine the influence of the initial values of the slip system strengths. 
Second, we consider the spatial arrangements of the phases and grains.  For this study, four types of microstructures are defined, ranging in degree of similarity that they bear with the actual duplex steel.    Following presentation of the two sub-studies we summarize the collective findings and offer concluding thoughts.

\clearpage

\section{Choice of Constitutive Parameters}
\label{sec:parameters}
There are several challenges inherent in characterizing the material parameters of dual phase system:
\begin{enumerate}
\item  the constituent phases cannot readily be physically separated and tested individually in a load frame; 
\item data pertaining to the mechanical responses of the individual phases are required, so there are twice as many parameters to optimize; and 
\item the parameters cannot be determined from macroscopic data alone. 
\end{enumerate}
In the face of these challenges, a strategy to identify the  
material parameters from a combination of multiple references and new experiments
was employed. Tabulated values were used for the single crystal elastic constants. Uniaxial strain rate jump tests were employed to characterize the rate sensitivity. Macroscopic stress-strain data for monotonic uniaxial loading at constant strain rate were used to characterize the remaining plasticity parameters, assuming equal values for both phases. 
These data included lattice strain measurements, acquired using neutron diffraction.
It should be emphasized that only the uniaxial lattice strain data were used for fitting the material parameters. 
The other four sets of lattice strain data, corresponding to biaxial loading, had no role in the parameter selection. 

This approach to fitting material parameters is similar to the methodology developed by Baczmanski and Braham~\cite{Baczmanski04a} for duplex stainless steel. They only compared lattice strains for one crystallographic fiber per phase. However, the lattice strain response of one fiber is not representative of the entire phase. Wong and Dawson~\cite{Wong10a} demonstrated that for even single phase materials, the location and nature of inflection points differ from fiber to fiber and depend on directional strength-to-stiffness. It is therefore critical to compare experimental and simulated lattice strain responses across several fibers per phase, as was done by Baczmanski, Dakhlaoui, and colleagues in subsequent works~\cite{Dakhlaoui06a, Dakhlaoui07a}.

\begin{figure}[h]
	\centering
	\subfigure[Grain]{\includegraphics[width=0.49\linewidth]{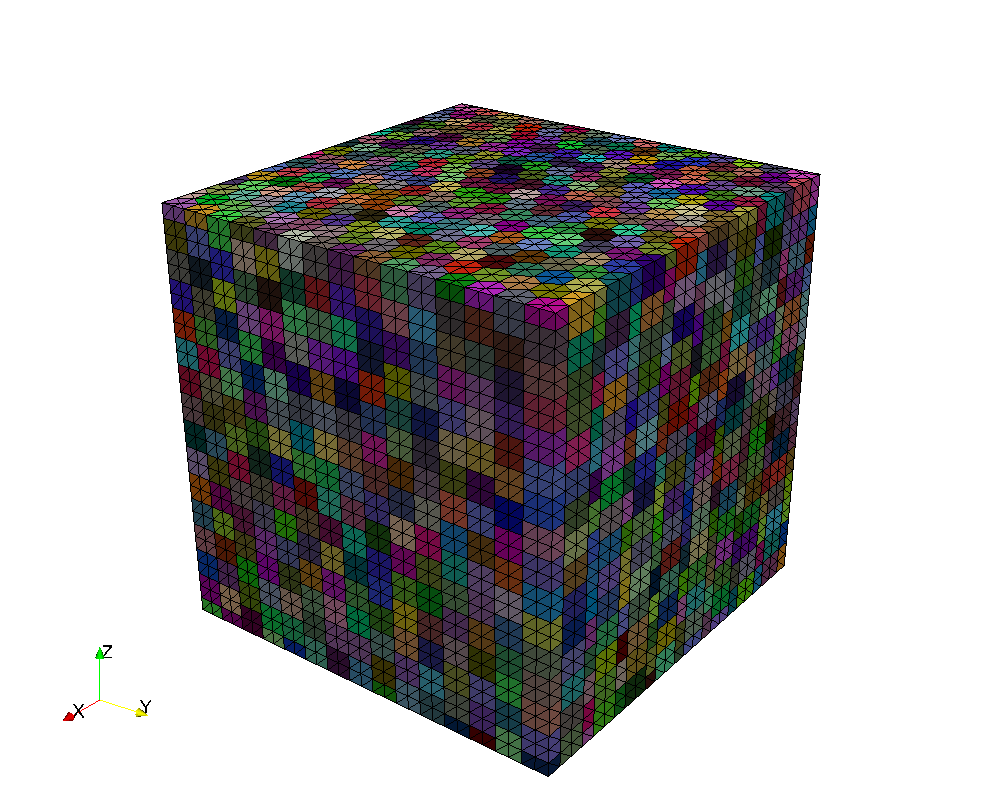}\label{fig:equiaxed-columnar-grain-1}}
	\subfigure[Phase]{\includegraphics[width=0.49\linewidth]{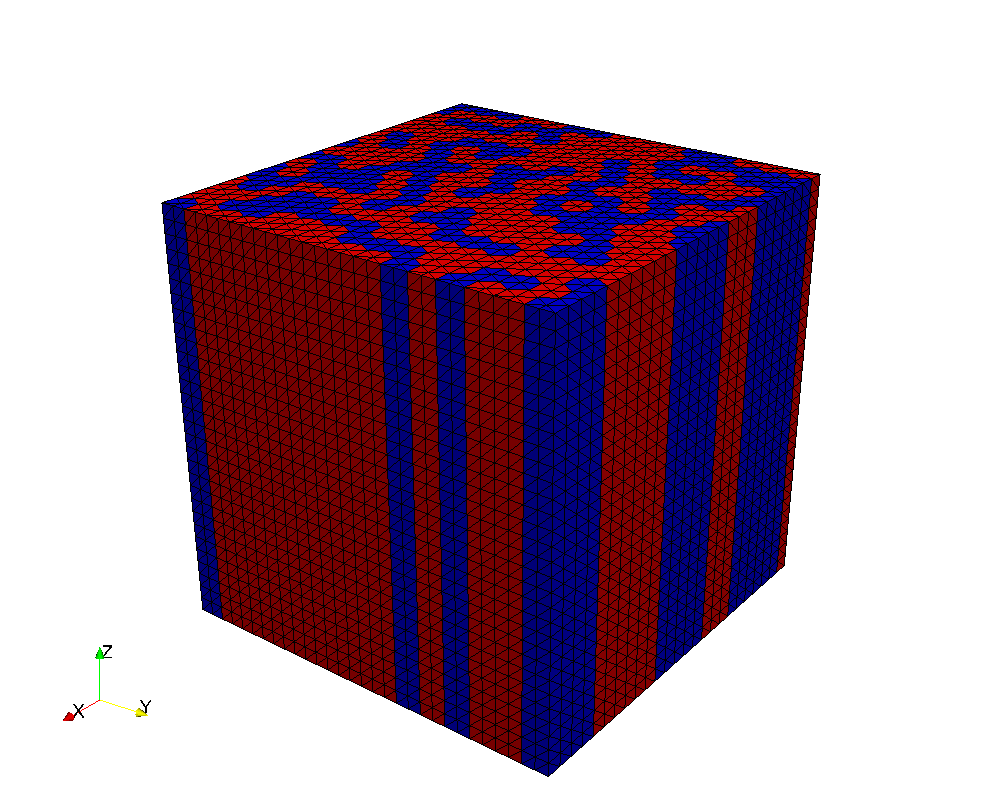}\label{fig:equiaxed-columnar-phase-1}}
	\caption{Finite element mesh of the equiaxed hexagonal grain, columnar phase microstructure used in the simulations for determination of the constitutive model parameters.}
	\label{fig:mesh-equiaxed-columnar_Const-Parms}
\end{figure}

An overview of the modeling of the {\it in situ} loading diffraction experiments is given in the companion paper~\cite{pos_daw_twophase}.
Virtual specimens are needed in the modeling.  Section~\ref{sec:microstructure} details the effort devoted to simulating different virtual specimens and assessing the sensitivity of the predicted lattice strains to attributes of the specimens.
The simulations to determine the constitutive model parameters, which is the focus of this section of the paper, were conducted with one particular virtual specimen.   
This sample had  an equiaxed columnar microstructure with  
2,320 austenite grains and 1,753 ferrite grains. 
The grains of the microstructure are discretized with a total of 137,700 10-node, tetrahedral finite elements.  
The mesh is shown on backgrounds that shown the grains (Part a) and the phases (Part b) in Figure~\ref{fig:mesh-equiaxed-columnar_Const-Parms}.

\subsection{Constitutive model for elastoplastic behavior}\label{sec:model_formulation}

A brief overview of the constitutive model is presented. For more detailed information the reader is referred to~\cite{Dawson14a}.
The constitutive model employs a kinematic decomposition of the deformation into a sequence of deformations due to crystallographic slip, rotation, and elastic stretch. Using this decomposition, the deformation gradient, $\boldsymbol{F}$, can be represented as
\begin{equation}\label{eqn:kinematic_decomp}
\boldsymbol{F} = \boldsymbol{V}^e  \boldsymbol{R}^*  \boldsymbol{F}^p 
\end{equation}
where $\boldsymbol{F}^p$, $\boldsymbol{R}^*$, and  $\boldsymbol{V}^e$ correspond to crystallographic slip, rotation, and elastic stretch, respectively. A schematic of this decomposition is provided in~\cite{Dawson14a}. This decomposition defines a reference configuration $\mathscr{B}_0$, a deformed configuration $\mathscr{B}$, and two intermediate configurations $\bar{\mathscr{B}}$ and $\hat{\mathscr{B}}$. The state equations are written in the intermediate $\hat{\mathscr{B}}$ configuration defined by the relaxation of the elastic deformation from the current $\mathscr{B}$ configuration. Elastic strains, $\boldsymbol{\epsilon}^e$, are required to be small, which allows the elastic stretch tensor to be written as
\begin{equation}
\boldsymbol{V}^e = \boldsymbol{I} + \boldsymbol{\epsilon}^e
\end{equation}
The velocity gradient, $\boldsymbol{L}$, is calculated from the deformation gradient using the relationship
\begin{equation}
\boldsymbol{L} = \dot{\boldsymbol{F}} \boldsymbol{F}^{-1}
\end{equation}
The velocity gradient can be decomposed into a symmetric deformation rate, $\boldsymbol{D}$, and skew-symmetric spin rate, $\boldsymbol{W}$
\begin{equation}
\boldsymbol{L} = \boldsymbol{D} + \boldsymbol{W}
\end{equation}
A generic symmetric tensor, $\boldsymbol{A}$, can be additively decomposed into mean and deviatoric components
\begin{equation}\label{eqn:vol_dev_decomp}
\boldsymbol{A} = \sfrac{1}{3} \mathrm{tr} (\boldsymbol{A}) \boldsymbol{I} + \boldsymbol{A}^\prime
\end{equation}
where $\sfrac{1}{3} \mathrm{tr} (\boldsymbol{A})$ is the scalar mean component and $\boldsymbol{A}^\prime$ is the tensorial deviatoric component.
Utilizing Equations~\ref{eqn:kinematic_decomp}-\ref{eqn:vol_dev_decomp} the volumetric deformation rate, deviatoric deformation rate, and spin rate can be expressed as
\begin{equation}
\mathrm{tr} \left( \boldsymbol{D} \right) = \mathrm{tr} \left( \dot{\boldsymbol{\epsilon}}^e \right)
\label{eq:trD}
\end{equation}
\begin{equation}
\boldsymbol{D}^\prime = \dot{\boldsymbol{\epsilon}}^{e\prime} + \hat{\boldsymbol{D}}^p + \boldsymbol{\epsilon}^{e\prime} \cdot \hat{\boldsymbol{W}}^p - \hat{\boldsymbol{W}}^p \cdot \boldsymbol{\epsilon}^{e\prime}
\end{equation}
\begin{equation}\label{eqn:spin_rate}
\boldsymbol{W} = \hat{\boldsymbol{W}}^p + \boldsymbol{\epsilon}^{e\prime} \cdot \hat{\boldsymbol{D}}^p - \hat{\boldsymbol{D}}^p \cdot \boldsymbol{\epsilon}^{e\prime}
\end{equation}
where $\hat{\boldsymbol{D}}^p$ and $\hat{\boldsymbol{W}}^p$ are the plastic deformation and spin rates.

Constitutive equations relate the stress to the deformation. The Kirchhoff stress, $\boldsymbol{\tau}$, in the $\hat{\mathscr{B}}$ configuration is related to the Cauchy stress, $\boldsymbol{\sigma}$, in the current configuration $\mathscr{B}$ by the determinant of the elastic stretch tensor
\begin{equation}
\boldsymbol{\tau} = \mathrm{det} (\boldsymbol{V^e}) \boldsymbol{\sigma}
\end{equation}
The Kirchhoff stress is related to the elastic strain through anisotropic Hooke's law
\begin{equation}
\mathrm{tr} \left( \boldsymbol{\tau} \right) = 3K \mathrm{tr} \left( \boldsymbol{\epsilon}^e \right)
\label{eq:bulk-mod}
\end{equation}
\begin{equation}
\boldsymbol{\tau}^\prime = \mathscr{C}^* \boldsymbol{\epsilon}^{e\prime}
\label{eq:shear-mod}
\end{equation}
where $K$ is the bulk modulus and $\mathscr{C}^*$ is the fourth-order stiffness tensor.

Plastic deformation due to crystallographic slip occurs on a restricted set of slip systems. For FCC crystals, slip occurs on the \{111\} planes in the [110] directions. For BCC crystals, slip on the \{110\} planes in the [111] directions is considered. The plastic deformation rate and plastic spin rate are given by 
\begin{equation}
\hat{\boldsymbol{D}}^p = \sum_\alpha \dot{\gamma}^\alpha \hat{\boldsymbol{P}}^\alpha
\end{equation}
\begin{equation}
\hat{\boldsymbol{W}}^p = \dot{\boldsymbol{R}^*} \boldsymbol{R}^{*T} + \sum_\alpha \dot{\gamma}^\alpha \hat{\boldsymbol{Q}}^\alpha
\end{equation}
where $\hat{\boldsymbol{P}}^\alpha$ and $\hat{\boldsymbol{Q}}^\alpha$ are the symmetric and skew symmetric components of the Schmid tensor $\hat{\boldsymbol{T}}^\alpha$, and $\dot{\gamma}^\alpha$ is the shear rate on the $\alpha$-slip system. The Schmid tensor is defined as the dyad of the slip direction, $\hat{\mathbf{s}}^\alpha$, and slip plane normal, $\hat{\mathbf{m}}^\alpha$
\begin{equation}
\hat{\boldsymbol{T}}^\alpha = \hat{\mathbf{s}}^\alpha \otimes \hat{\mathbf{m}}^\alpha
\end{equation}
The slip system shear rate for a given slip system is related to the critical resolved shear stress on that slip system, $\tau^\alpha$, by a power law relationship
\begin{equation}
\dot{\gamma}^\alpha = \dot{\gamma}_0 \left( \frac{\vert \tau^\alpha \vert}{g^\alpha} \right)^\frac{1}{m} \mathrm{sgn} \left( \tau^\alpha \right)
\label{eq:fixed-state-kinetics}
\end{equation}
where $\dot{\gamma}_0$ is a reference slip system shear rate and $g^\alpha$ is the slip system strength. The resolved shear stress is the projection of the deviatoric stress onto the slip system
\begin{equation}
\tau^\alpha = \boldsymbol{\tau}^\prime : \hat{\boldsymbol{P}}^\alpha
\label{eq:schmidtensor}
\end{equation}

A modified Voce hardening law is used to describe slip system strength evolution 
\begin{equation} \label{eqn:Voce}
\dot{g}^\alpha = h_0 \left( \frac{g_s - g^\alpha}{g_s - g_0} \right)^{n^\prime} \sum_\alpha \dot{\gamma}^\alpha
\end{equation}
where $h_0$ is the reference hardening rate, $g_s$ is the saturation strength, and $n^\prime$ is the hardening exponent. The hardening law is isotropic; at a given material point, all slip systems harden at the same rate. Equation~\ref{eqn:spin_rate} describes the crystal reorientation.

An implicit time integration scheme is used to solve for the velocity field and material state.  An estimated material state at the end of each time increment is used to evaluate the constitutive equations. Implicit integration ensures stability. The solution algorithm for each time increment begins with initializing an initial guess of the velocity field. The deformed geometry at the end of the time increment is then estimated, based on the velocity field. The velocity gradient is computed and used to solve for the crystal state ($\mathrm{tr} (\boldsymbol{\epsilon}^e)$, $\boldsymbol{\epsilon}^{e\prime}$, $\boldsymbol{R}^*$, and $g^\alpha$) at each quadrature point. Constitutive matrices for the equilibrium equation (see Section~\ref{sec:microstructure}) are computed using the updated material state and used to solve for an updated velocity field. Iteration continues on the geometry, crystal state, and velocity field until the velocity solution is converged. The solution then advances to the next time increment.

\subsection{Elastic parameter identification}\label{sec:elastic_parm_selection}

Single crystal elastic constants appearing in Hooke's Law (Equations~\ref{eq:bulk-mod} and  \ref{eq:shear-mod}) were procured from archival literature. Ledbetter's elastic constants for austenitic stainless steel~\cite{Ledbetter01a} were used for austenite and Simmons and Wang's constants for pure BCC iron were used for ferrite~\cite{Simmons71a}. While alloying affects elastic moduli, these values provided reasonable estimates of the elastic constants for LDX-2101. These constants produced a good match between simulation and experiment for both fiber-averaged lattice strains in the elastic region and the macroscopic Young's modulus. 

\subsection{Plastic parameter identification}\label{sec:plastic_parm_selection}

The remainder of this section describes the iterative procedure used to obtain the plasticity parameters. 
In particular, the values of  $m$, $n^\prime$, $\dot{\gamma}_0$, $h_0$, $g_0$ and $g_s$ must be
evaluated.  
 The rate sensitivity, $m$ appears in the fixed-state kinetics equation, Equation~\ref{eq:fixed-state-kinetics},
 while the remaining parameters,   $n^\prime$, $\dot{\gamma}_0$, $h_0$, $g_0$ and  $g_s$, appear in the evolution equation for the slip system strengths, Equation~\ref{eqn:Voce}. 
The parameters for AL6XN~\cite{Marin12a}, adjusted for a reference strain rate of $\dot{\gamma}_0 = 10^{-4}$~s$^{-1}$, were used as an initial guess of the plasticity parameters for both phases.  The parameters were iteratively adjusted to match the macroscopic stress-strain curve.

\begin{figure}[h]
	\centering
	\subfigure[Jump test 1.]{\includegraphics[width=0.49\linewidth]{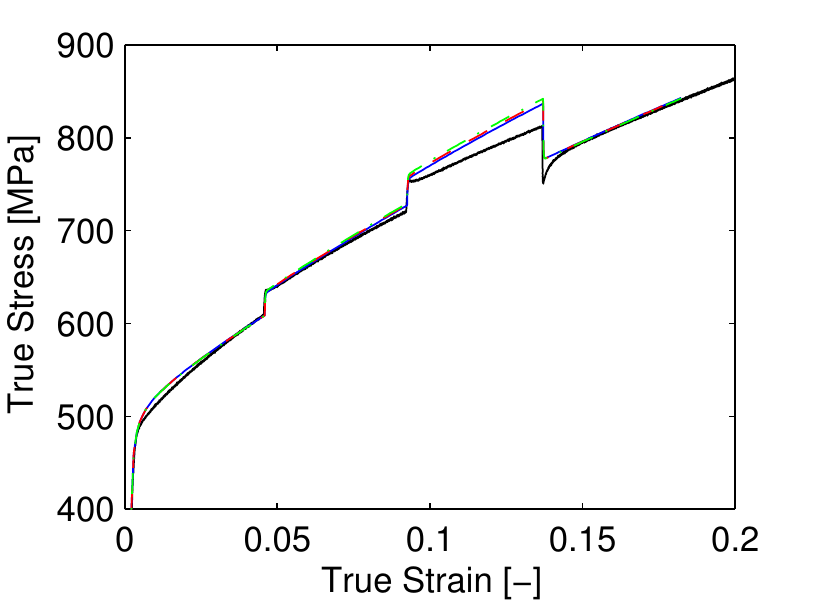}\label{fig:jump1}}
	\subfigure[Jump test 1 (zoom).]{\includegraphics[width=0.49\linewidth]{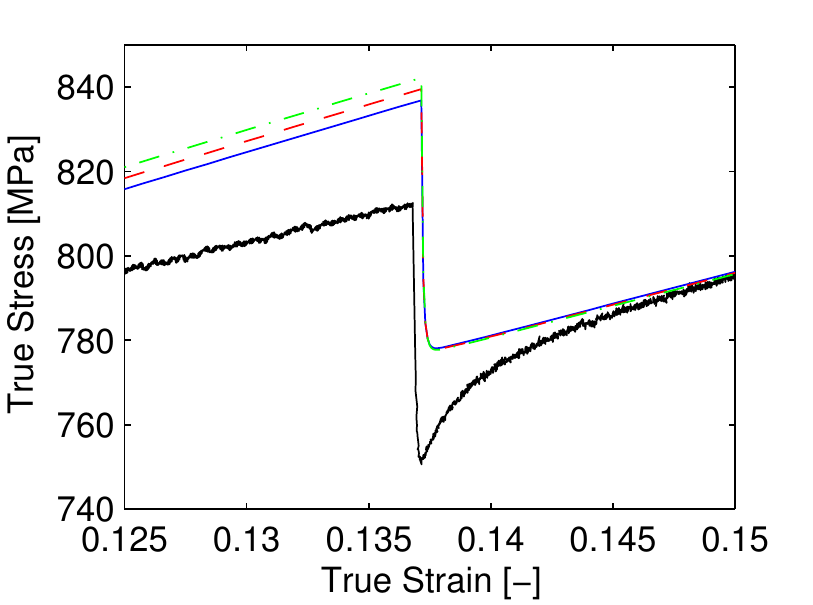}\label{fig:jump1zoom}}
	\subfigure[Jump test 2]{\includegraphics[width=0.49\linewidth]{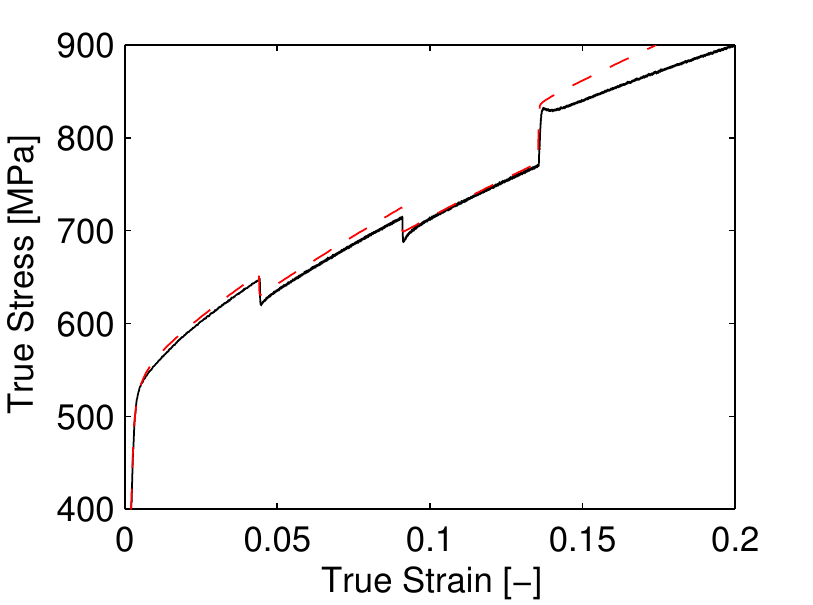}\label{fig:jump2}}
	\subfigure{\includegraphics[width=0.45\linewidth]{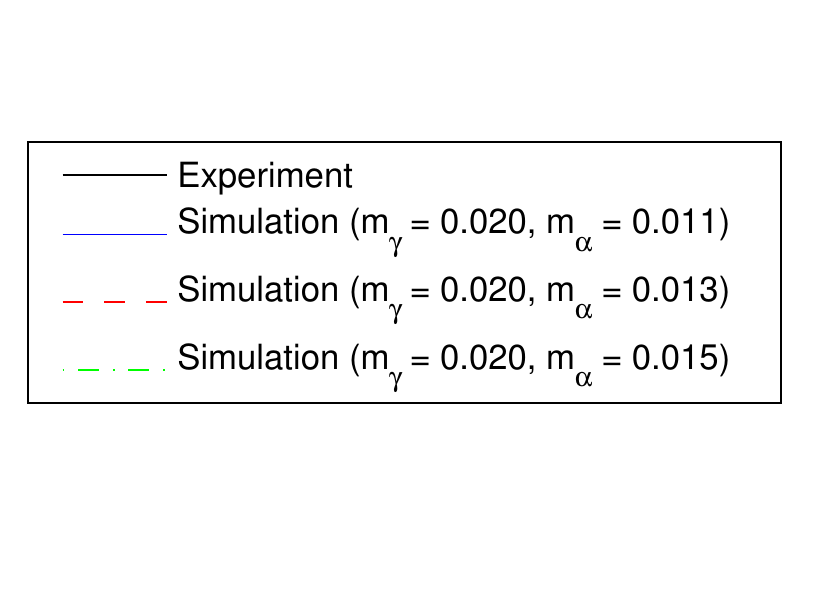}}
	\caption{Strain rate jump tests for LDX-2101 under uniaxial tension.}
	\label{fig:JumpTests}
\end{figure}

Strain rate jump tests were performed to calculate the rate sensitivity ($m$). In the first test (Figures~\ref{fig:jump1} and~\subref{fig:jump1zoom}), the nominal strain rate jumped from $10^{-4}$~s$^{-1}$, to $10^{-3}$~s$^{-1}$, then to $10^{-2}$~s$^{-1}$, and back to $10^{-4}$~s$^{-1}$. In the second test (Figure~\ref{fig:jump2}), the strain rate jumped from $10^{-2}$~s$^{-1}$, to $10^{-3}$~s$^{-1}$, then to $10^{-4}$~s$^{-1}$, and back to $10^{-2}$~s$^{-1}$. The combined macroscopic strain rate sensitivity is calculated from the changes in true stress and true strain rate at a jump
\begin{equation}\label{eqn:rate_sensitivity}
m = \frac{\partial \ln \sigma}{\partial \ln \dot{\epsilon}} \approx
\frac{\mathrm{ln}(\sigma_2) - \mathrm{ln}(\sigma_1)}{\mathrm{ln}(\dot{\epsilon}_2) - \mathrm{ln}(\dot{\epsilon}_1)}
\end{equation}
The experimental combined rate sensitivity is 0.017. The rate sensitivity of austenite was assumed to be 0.020, based on the value used by Marin et al.~\cite{Marin12a}. The rate sensitivity of ferrite was iteratively adjusted, until the simulated combined rate sensitivity matched that of the experiment. Based on this analysis, the rate sensitivity of ferrite is 0.013. These values agree with Talyan et al.~\cite{Talyan98a}, who state that austenite is more rate sensitive that ferrite and that the rate sensitivity of austenite is between 0.015 and 0.020. 
%

\begin{figure}[h]
	\centering
	\subfigure[Full.]{\includegraphics{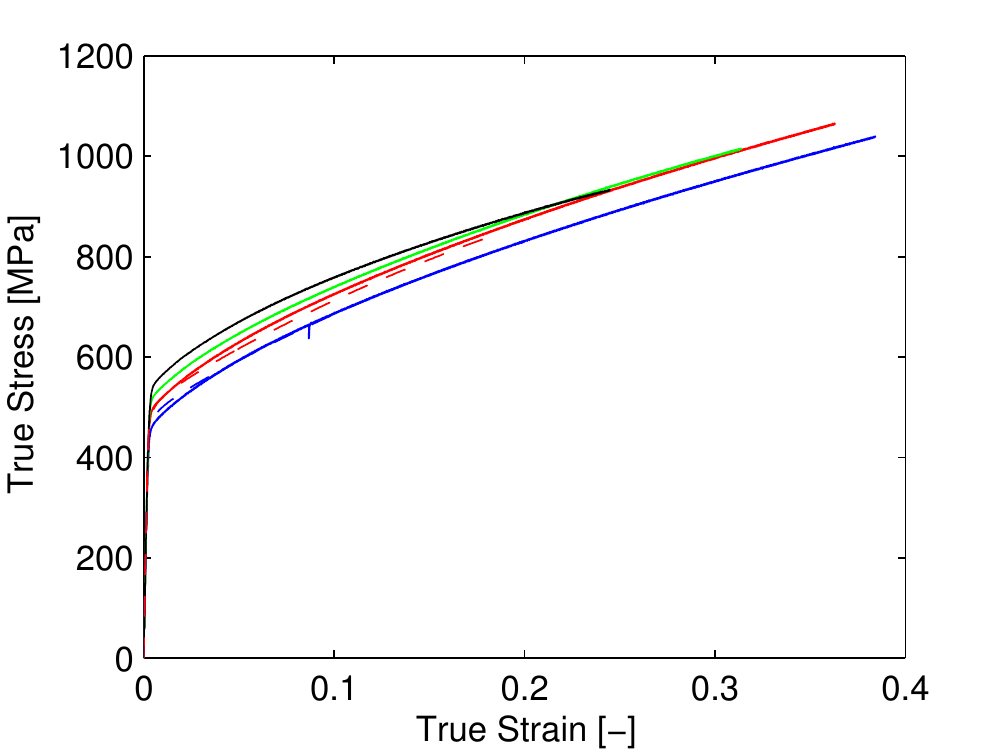}\label{fig:SigEpsMono}}
	\subfigure[Zoom.]{\includegraphics{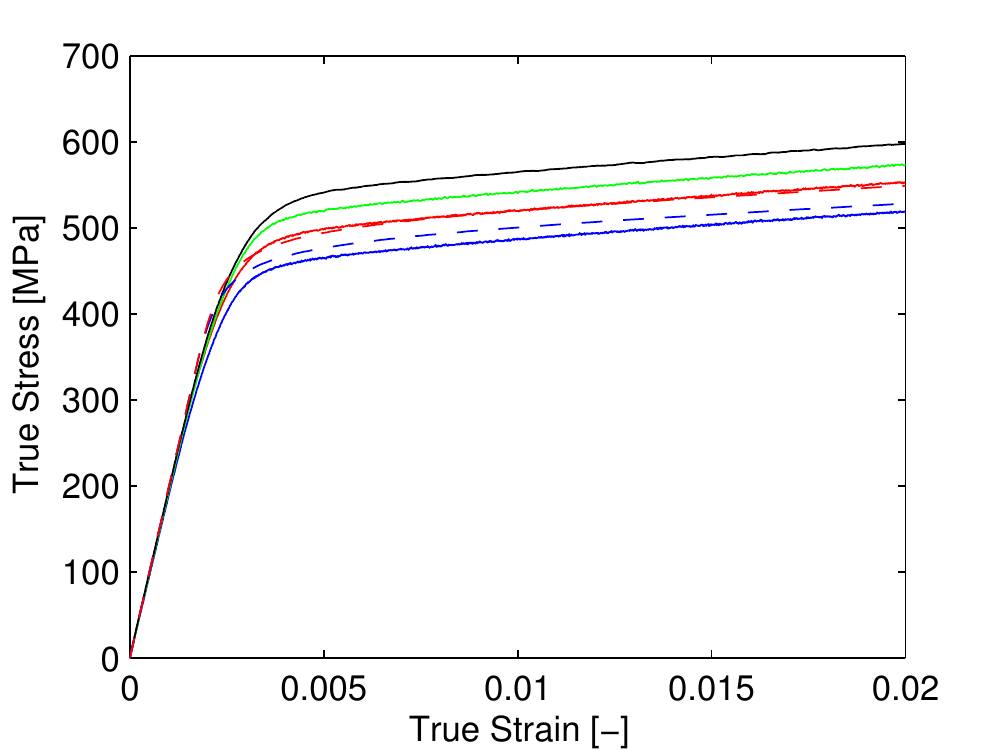}\label{fig:SigEpsMonoZoom}}
	\fbox{
	\subfigure{\includegraphics[trim = 0.2in 1.1in 0.2in 0.2in, clip]{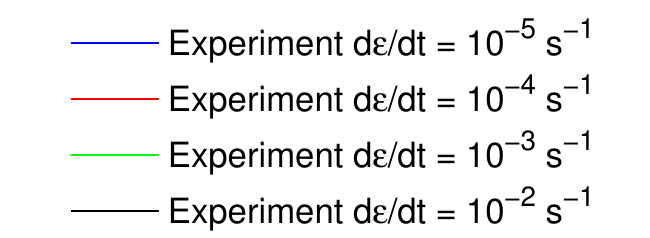}}
	\subfigure{\includegraphics[trim = 0.2in 1.1in 0.2in 0.2in, clip]{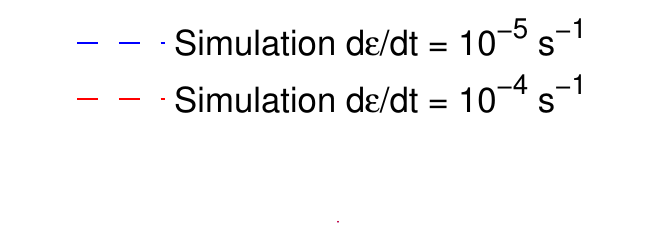}}}
	\caption{Monotonic true stress-strain responses of LDX-2101 under uniaxial tension.}
	\label{fig:StressStrain}
\end{figure}

\begin{figure}[h]
\centering
\includegraphics[width=0.8\linewidth]{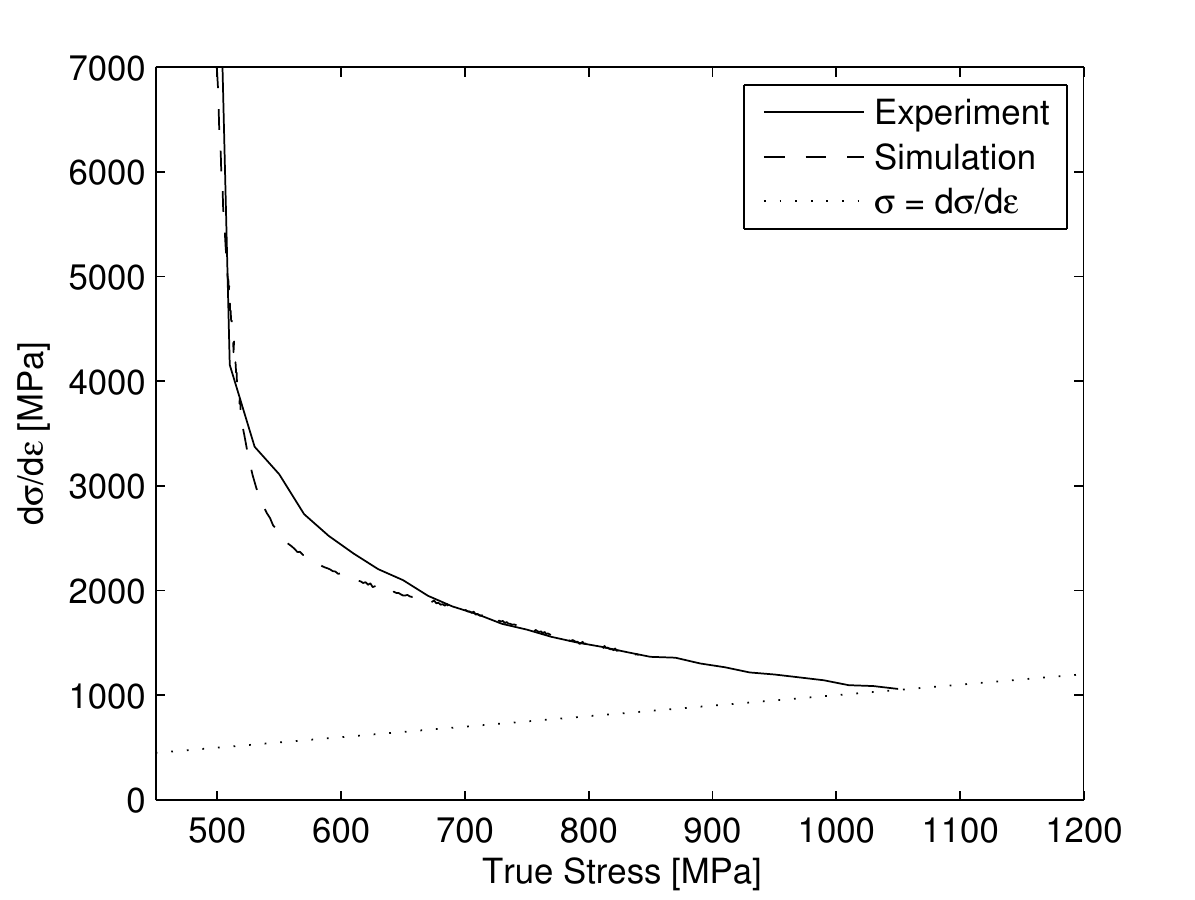}
\caption{Macroscopic hardening rate as a function of macroscopic stress ($d\epsilon/dt = 10^{-4}~\mathrm{s}^{-1}$).}\label{fig:dsigdeps}
\end{figure}

The macroscopic stress-strain curves for uniaxial monotonic loading were refit out to 20$\%$ strain, following the determination of the rate sensitivities (Figure~\ref{fig:StressStrain}). The simulation captures the increase in yield stress associated with an increase in strain rate. The exponent ($n^\prime$) in the hardening law can be obtained by fitting the hardening curve (Figure~\ref{fig:dsigdeps}). Given the already large material parameter space, a value of unity was chosen for simplicity. This value provides reasonable agreement between the simulated and experimental hardening curves, although there is room for additional improvement. One method for improving the hardening model would be to add a term to Equation \ref{eqn:Voce} that scales with the slip system shear rates $h_1 \sum_\alpha \dot{\gamma}^\alpha$. The additional term would capture the hardening that occurs during fully-developed plasticity. Then $h_0$ and $n^\prime$ could be adjusted to better model hardening during the elasto-plastic transition.

The final step of the material parameter fitting was to differentiate the hardening coefficients ($h_0$), initial slip system strengths ($g_0$), and saturation slip system strengths ($g_s$) for each phase. The relative strengths of austenite and ferrite vary greatly among duplex stainless steel alloys. For some alloys, ferrite is stronger than austenite, and for others, the two phases are of similar strength. Relative strength is dependent of chemical composition. Alloying with nitrogen, in particular, preferentially hardens the austenitic phase and can produce alloys in which austenite is stronger than ferrite~\cite{Foct93a}. Microhardness data, while not linearly related to slip system strength, provide an indication of relative phase strength. Microhardness data suggest that for alloys UR35N and UR52N+, ferrite is stronger than austenite~\cite{Dakhlaoui07a}, whereas for SAF 2304, austenite is stronger~\cite{Johansson99a}.

The plasticity parameters employed by other researchers for various duplex stainless steel alloys serve as benchmarks for LDX-2101. Tables~\ref{tab:ss_strength} and~\ref{tab:hard_coeff} summarize values of slip system strength and hardening coefficient found in the literature. Baczmanski and Braham~\cite{Baczmanski04a}, Dakhlaoui et al.~\cite{Dakhlaoui06a, Dakhlaoui07a}, and Jia et al.~\cite{Jia08a} differentiated the plasticity parameters for the two phases by fitting lattice strains from elasto-viscoplastic self-consistent models to neutron diffraction measurements for uniaxial loading. For the alloys studied, they report that the slip system strength of ferrite is between 1.6 and 3.3 times greater than that of austenite. These researchers all considered slip on both the \{110\}~[111] and \{211\}~[111] slip systems for ferrite. Considering slip on just the \{110\}~[111] slip systems for ferrite reduced the ratio of slip system strengths, obtained by fitting the lattice strain data, by only six percent~\cite{Baczmanski04a}. Hedstrom et al.~\cite{Hedstrom10a} fit material parameters using macroscopic data from single-phase austenite and ferrite. They report that the slip system strengths for the two phases are roughly equal with austenite stronger than ferrite. Chemical composition and processing both affect the plasticity parameters. It is therefore questionable whether the single-phase materials used to fit the plasticity parameters had similar composition and processing as the duplex material. However, Hedstrom's study reveals that it is possible for austenite and ferrite to have similar slip system strengths. In all studies, austenite is shown to harden faster than ferrite. In most cases, the hardening coefficient for austenite is between 1.5 and 2.7 times greater than that of ferrite.

\begin{table}[h]
	\centering
	\caption{Slip system strengths for duplex stainless steel alloys. The ratio between BCC \{110\} [111] and FCC \{111\} [110] slip system strength is also tabulated.}	
	\begin{tabular} {c c c c c c }
		& & \multicolumn{3}{c}{Initial slip system strength (MPa)} & \\
		& & FCC & BCC & BCC & \\
		Reference & Alloy & \{111\} [110] & \{110\} [111] & \{110\} [211] & Ratio \\ \hline
		Baczmanski~\cite{Baczmanski04a} & & 75 & 230 & - & 3.1 \\
		Baczmanski~\cite{Baczmanski04a} & & 75 & 245 & 245 & 3.3 \\
		Dakhlaoui~\cite{Dakhlaoui06a} & UR45N & 140 & 220 & 220 & 1.6 \\
		Dakhlaoui~\cite{Dakhlaoui06a} & UR45N & 120 & 200 & 200 & 1.7 \\
		Dakhlaoui~\cite{Dakhlaoui07a} & UR35N & 70 & 190 & 190 & 2.7 \\
		Dakhlaoui~\cite{Dakhlaoui07a} & UR45N & 115 & 260 & 260 & 2.3 \\
		Dakhlaoui~\cite{Dakhlaoui07a} & UR52N+ & 150 & 330 & 330 & 2.2 \\
		Jia~\cite{Jia08a} & SAF 2507 & 98 & 230 & 250 & 2.3 \\
		Hedstrom~\cite{Hedstrom10a} & SAF 2304 & 137 & 111& ? &0.8
	\end{tabular}
	\label{tab:ss_strength}
\end{table}

\begin{table}[h]
	\centering
	\caption{Hardening coefficients for duplex stainless steel alloys. The ratio between FCC and BCC hardening coefficients is also tabulated.}	
	\begin{tabular} {c c c c c}
		& & \multicolumn{2}{c}{Hardening coefficient (MPa)} & \\
		Reference & Alloy & FCC & BCC & Ratio \\ \hline
		Baczmanski~\cite{Baczmanski04a} & &  190 & 50 & 3.8\\
		Baczmanski~\cite{Baczmanski04a} & & 190 & 80 & 2.4 \\
		Dakhlaoui~\cite{Dakhlaoui06a} & UR45N & 200 & 120 & 1.7 \\
		Dakhlaoui~\cite{Dakhlaoui07a} & UR35N &  200 & 80 & 2.5\\
		Dakhlaoui~\cite{Dakhlaoui07a} & UR45N &  215 & 80 & 2.7\\
		Dakhlaoui~\cite{Dakhlaoui07a} & UR52N+ & 200 & 80 & 2.5\\
		Hedstrom~\cite{Hedstrom10a} & SAF 2304 & 600 & 400 & 1.5
	\end{tabular}
	\label{tab:hard_coeff}
\end{table}

To differentiate the hardening coefficients ($h_0$), initial slip system strengths ($g_0$), and saturation slip system strengths ($g_s$) for each phase, uniaxial simulations were performed using different sets of parameters. The simulation results were then compared with experimental lattice strain data. The ratios between the austenitic phase and ferritic phase parameters were adjusted while holding the volume-averaged parameters constant. In this manner, the macroscopic stress-strain response was preserved. As LDX-2101 is rather ductile at room temperature, with over 40\% elongation achieved during tensile tests before necking, neither phase reaches its saturation strength in the range of interest, between 0 and 20\% strain. For this reason, the saturation strength ($g_s$) ratio is of secondary importance to the initial hardness ($g_0$) and hardening coefficient ($h_0$) ratios. The saturation strength ratio is therefore chosen based on the other parameters, such that both phases saturate at approximately the same macroscopic strain. Four candidate sets of material parameters were selected to investigate whether ferrite is stronger than austenite and whether austenite hardens faster than ferrite for LDX-2101. These parameters are presented in Table~\ref{tab:HardParamStudy}. Simulations were performed using a mesh with equiaxed grains and columnar phase structure for both monotonic and experimental load histories.

\begin{table}[h]
	\centering
	\caption{Values for material parameter study in MPa.}
	\begin{tabular} {c | c c c | c c c}	
		& \multicolumn{3}{c |}{Austenite} & \multicolumn{3}{c}{Ferrite} \\
		& $(h_0)_\gamma$ & $(g_0)_\gamma$ & $(g_s)_\gamma$ & $(h_0)_\alpha$ & $(g_0)_\alpha$ & $(g_s)_\alpha$ \\ \hline
		A & 336 & 192 & 458 & 336 & 192 & 458 \\
		B & 428 & 192 & 531 & 214 & 192 & 361 \\
		C & 336 & 134 & 400 & 336 & 269 & 535 \\
		D & 428 & 134 & 473 & 214 & 269 & 438
	\end{tabular}	
	\label{tab:HardParamStudy}
\end{table}

\begin{figure}[h]
	\centering
	\subfigure[Full.]{\includegraphics{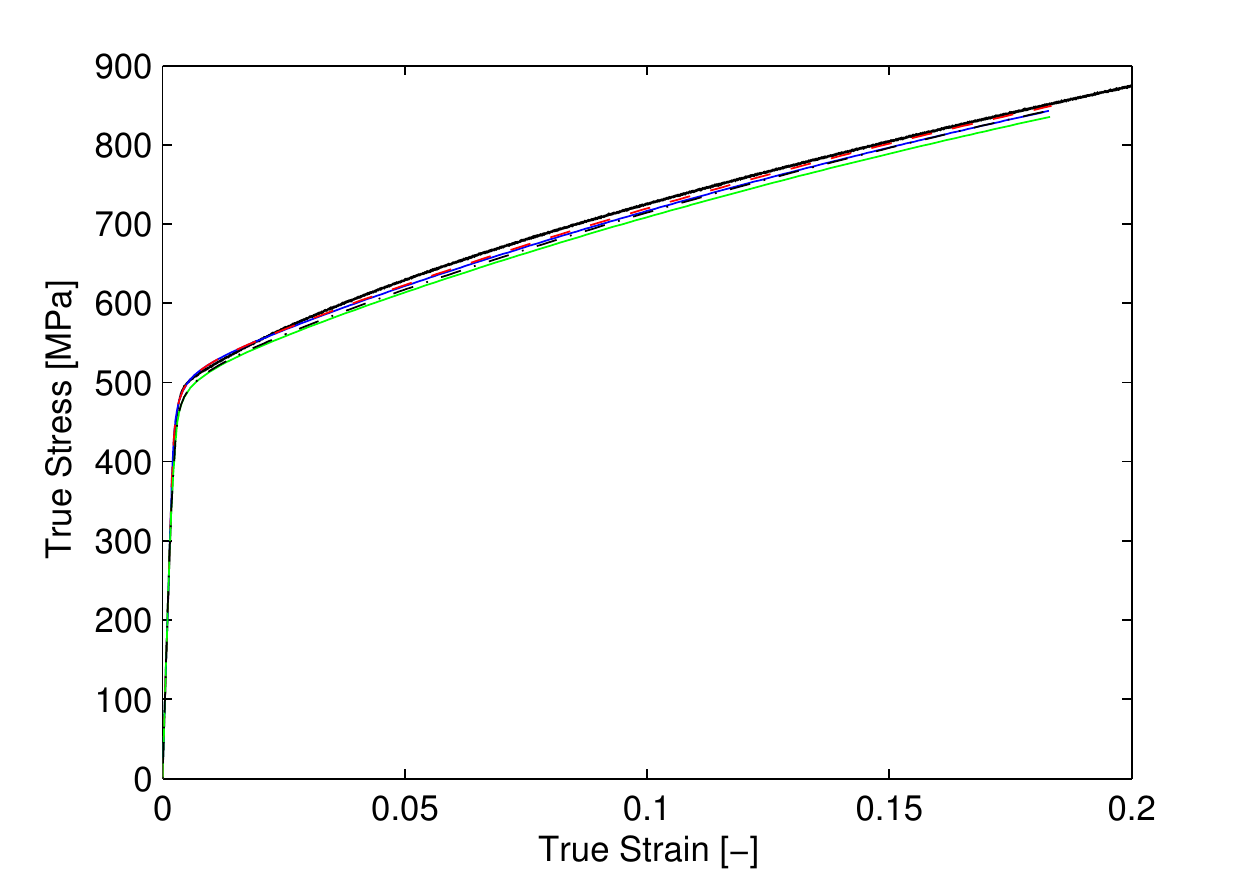}\label{fig:HardParamsStressStrainFull}}
	\subfigure[Zoom.]{\includegraphics{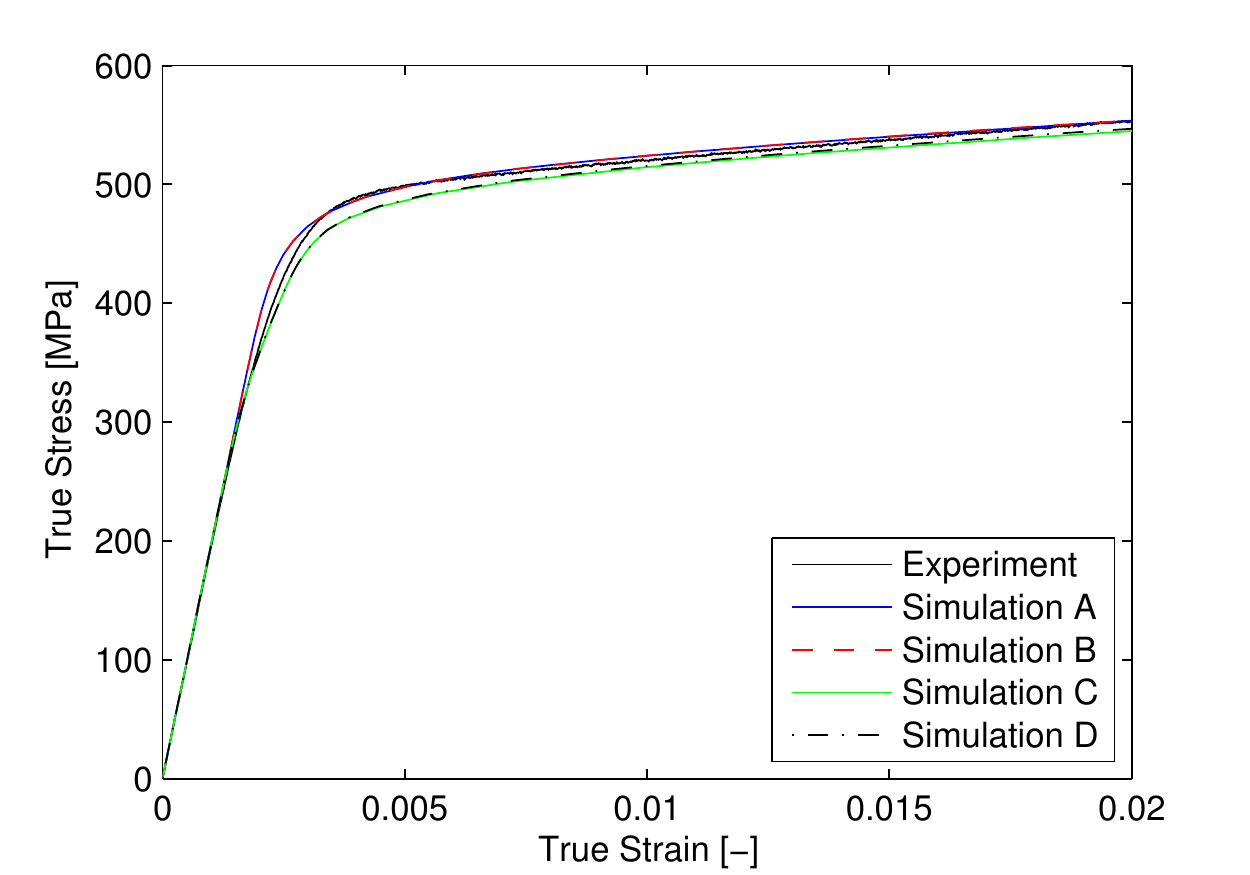}\label{fig:HardParamsStressStrainZoom}}
	\caption{Macroscopic stress-strain responses for the four sets of material parameters tabulated in Table~\ref{tab:HardParamStudy}}. 
	\label{fig:HardParamsStressStrain}
\end{figure}

The macroscopic stress-strain responses for monotonic loading out to 20\% strain are presented in Figure~\ref{fig:HardParamsStressStrainFull}. The macroscopic responses in the fully-developed plastic regime, while not identical, are relatively similar for the four sets of parameters. Examination of the knee of the stress-strain curve in Figure~\ref{fig:HardParamsStressStrainZoom} reveals differences in the response between the equal and unequal strength simulations. In the equal strength simulations, A and B, yielding occurs in both phases at around the same macroscopic stress. However, for Simulations C and D, in which ferrite is stronger than austenite, austenite yields at a lower macroscopic stress than ferrite, and the yielding is bimodal. As a result, the macroscopic stress-strain responses deviate from linearity at a lower macroscopic stress for the unequal strength simulations. While the unequal strength simulations exhibit better agreement with the experiment at small strain (around 0.002), the equal strength simulations  exhibit better agreement in the range of 0.003-0.010 strain. Neither pair of simulations does a substantially better job of modeling the macroscopic stress-strain response than the other pair.

\begin{figure}[h]
	\centering
	\subfigure[FCC \{200\}.]{\includegraphics[width=0.49\linewidth]{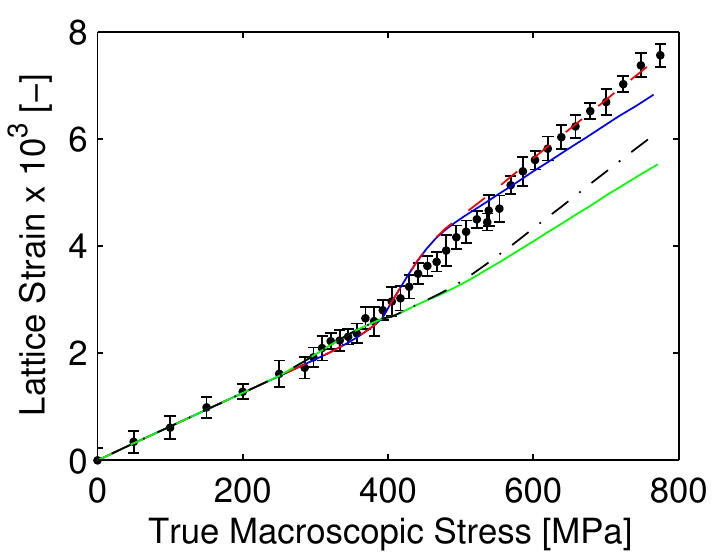}\label{fig:InitialHardnessFCC200}}
	\subfigure[BCC \{200\}.]{\includegraphics[width=0.49\linewidth]{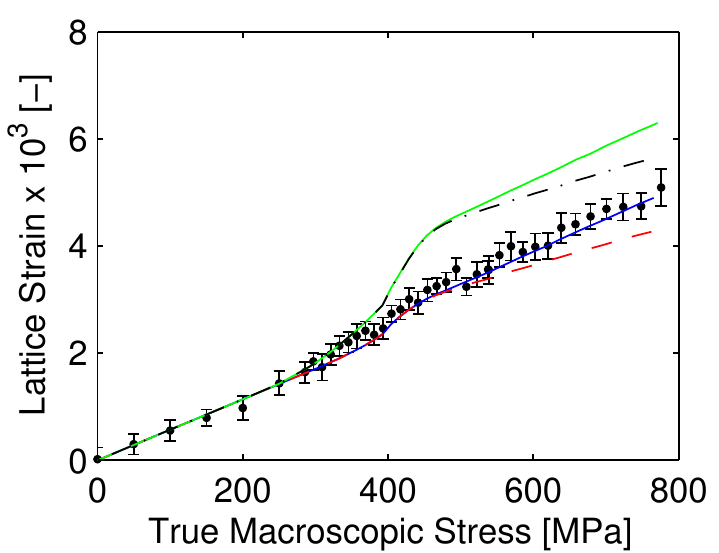}\label{fig:InitialHardnessBCC200}}
	\subfigure[FCC \{111\}.]{\includegraphics[width=0.49\linewidth]{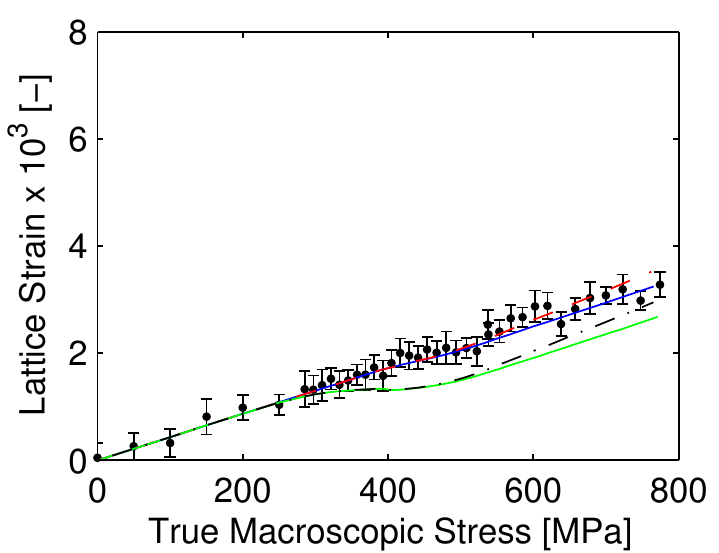}\label{fig:InitialHardnessFCC222}}
	\subfigure[BCC \{110\}.]{\includegraphics[width=0.49\linewidth]{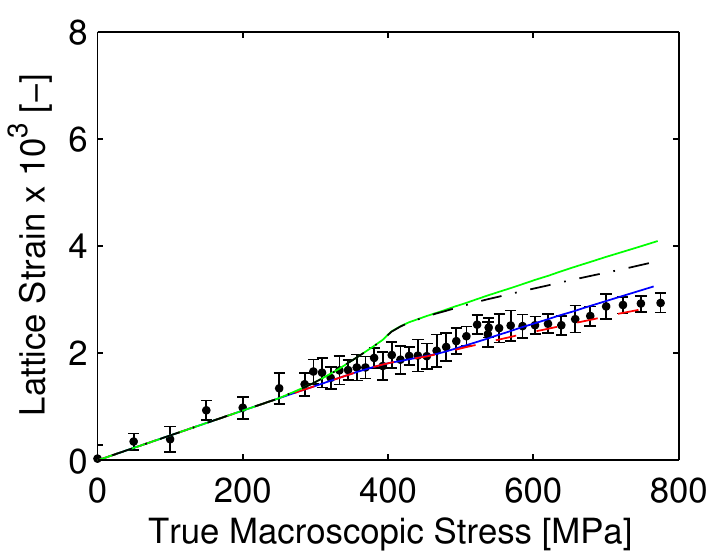}\label{fig:InitialHardnessBCC220}}
	\subfigure[FCC \{220\}.]{\includegraphics[width=0.49\linewidth]{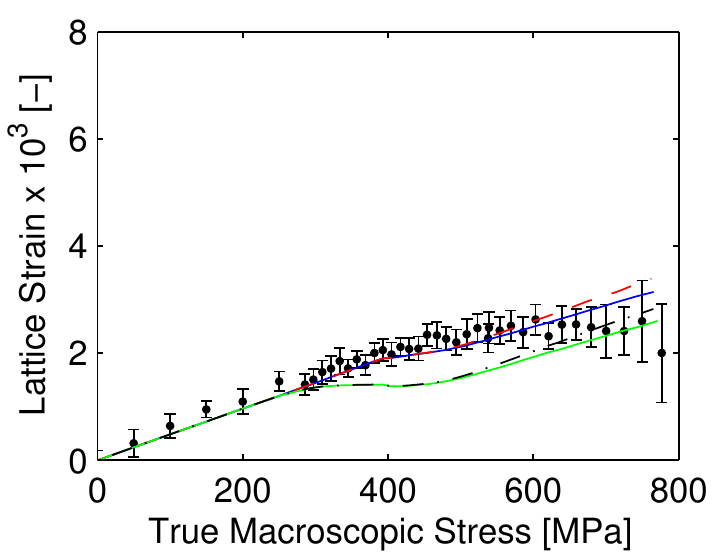}\label{fig:InitialHardnessFCC220}}
	\subfigure[BCC \{211\}.]{\includegraphics[width=0.49\linewidth]{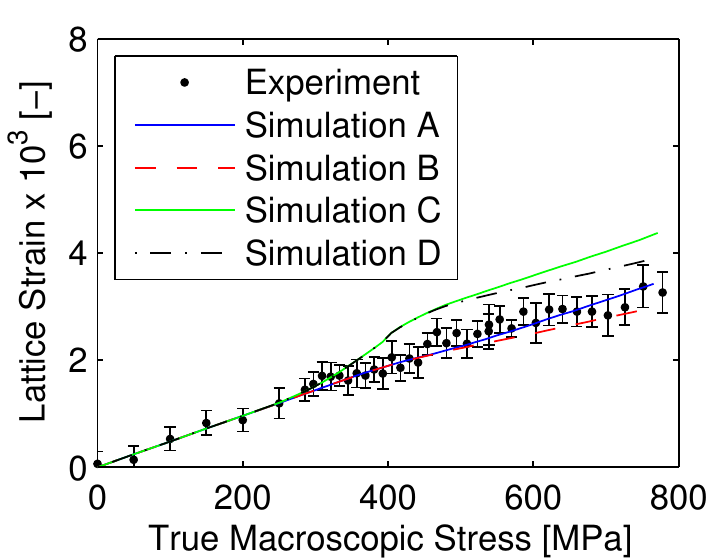}\label{fig:InitialHardnessBCC211}}
	\caption{Experimental and simulated fiber-averaged axial lattice strains for uniaxial loading. The simulation data correspond to different ratios of initial slip system strengths ($g_0$) and hardening coefficients ($h_0$) between the two phases. Material parameters for the simulations are tabulated in Table~\ref{tab:HardParamStudy}}
	\label{fig:LSHardParams}
\end{figure}

The differences in mechanical behavior of the two phases are revealed in the fiber-averaged lattice strain responses. Axial lattice strain for the experiment and simulations, conducted using the experimental load history, are presented in Figure~\ref{fig:LSHardParams}. The equal strength simulation results (A and B) agree with the experimental data, whereas the unequal strength simulation results (C and D) do not. In particular, the equal strength simulations capture the initial inflections that occur at the start of the elasto-plastic transition around 300 MPa. The nature of these inflections is governed almost exclusively by elastic anisotropy, which is known, and the initial slip system strengths. The difference in relative slip system strength between LDX-2101 and other duplex stainless steel alloys~\cite{Baczmanski04a, Dakhlaoui06a, Dakhlaoui07a, Jia08a} is evident in these inflections. For example, in Dakhlaoui et al.'s experimental data, all ferrite axial lattice strains exhibit initial upward inflection and all the austenite axial lattices strains, with the exception of the \{200\} fiber, exhibit clear initial downward curvature. This type of lattice strain signature points to ferrite being stronger than austenite. In contrast, for the LDX-2101 experimental data, the BCC \{110\} and \{211\} fibers exhibit downward inflections, and the FCC \{111\} and \{220\} fibers exhibit near linear behavior. Neither phase is strictly concave up nor concave down during the elasto-plastic transition. Since both phases have relatively similar stiffness, this behavior is indicative of the phases being of relatively equal strength.

The hardening coefficient ($h_0$) affects the slopes of the lattice strain curves in the fully-developed plastic regime between 500 and 800 MPa. It is not apparent from the lattice strain data whether austenite hardens faster or at the same rate as ferrite. For the FCC \{200\} fiber, Simulation B, in which austenite hardens faster, exhibits better agreement with the experimental data than Simulation A, in which both phases harden at the same rate. However, the opposite is true for the BCC \{200\} fiber, for which Simulation A, with equal hardening rates, exhibits better agreement. Both Simulations A and B exhibit similar levels of agreement with the experimental data for the other four fibers. Because the lattice strain data do not provide compelling evidence for different hardening coefficients for the two phases, the hardening coefficients were chosen to be equal.

Additional simulations were conducted with austenite stronger than ferrite by 10\% and ferrite stronger than austenite by 10\% to further refine the slip system strengths for each phase. These changes to the initial slip system strength did not significantly affect the lattice strain agreement, and so the initial slip system strength was therefore chosen to be equal for each phase.
The converged set of parameters are given in Section~\ref{sec:discussion}.

\clearpage

\section{Instantiation of Virtual Polycrystals}
\label{sec:microstructure}

LDX-2101 microstructure presents several challenges for the defining the phases and grains within them as well as for the subsequent mesh generation. The first challenge is the irregular shapes of the ferrite grains.  The second challenge to microstructure modeling is the disparate grain size between the two phases. Austenite grains are smaller, on the order of 20-50~$\mu$m diameter, whereas the larger ferrite grains can extend over 100~$\mu$m. The disparate grain sizes poses challenges for mesh refinement. Meshes were constructed by tessellating a hexagonal prismatic base mesh, comprised of multiple tetrahedral elements. The hexagonal prismatic base regions formed building blocks that were then stitched together to form grains. Annealing twins are just on the order of mesh resolution and were not represented in the virtual microstructure.

\begin{figure}[h]
	\centering
	\subfigure[Three-dimensional base structure.]{\includegraphics[width=0.49\linewidth]{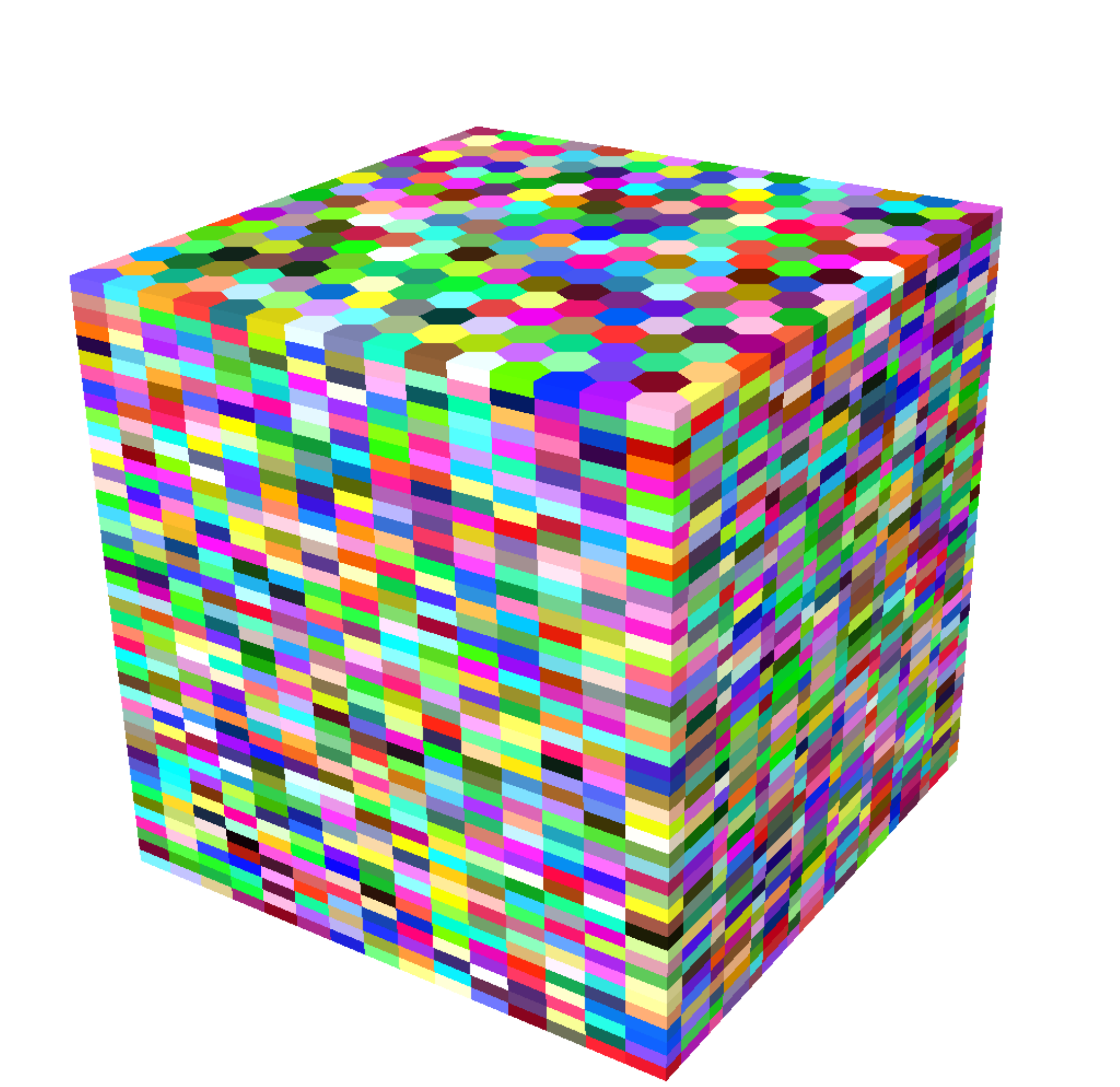}\label{fig:base_mesh}}
	\subfigure[Two-dimensional parent microstructure.]{\includegraphics[width=0.49\linewidth]{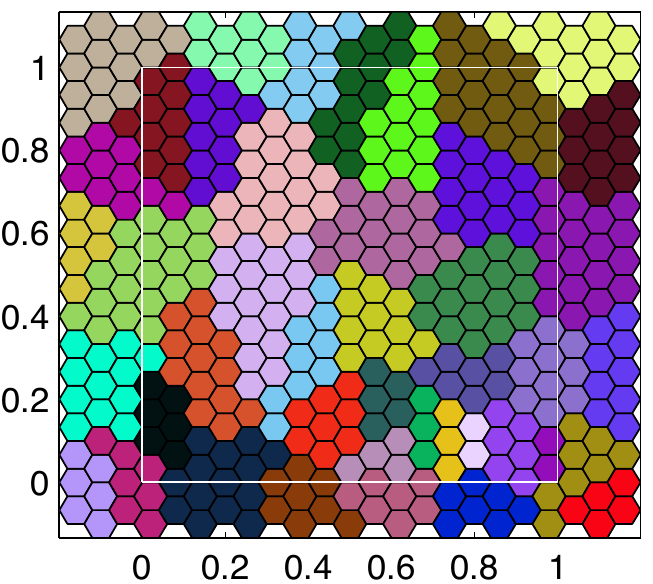}\label{fig:ferrite2D}}
	\subfigure[Two-dimensional microstructure. Austenite regions are colored white.]{\includegraphics[width=0.49\linewidth]{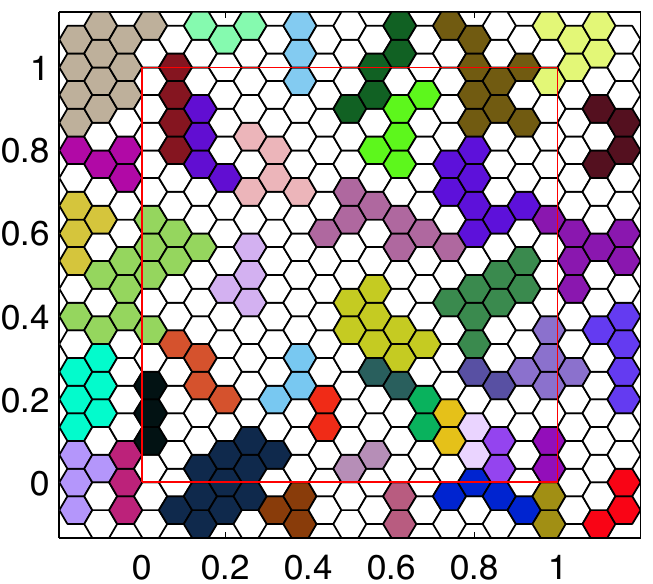}\label{fig:microstruct2D}}
	\subfigure[Three-dimensional microstructure.]{\includegraphics[width=0.49\linewidth]{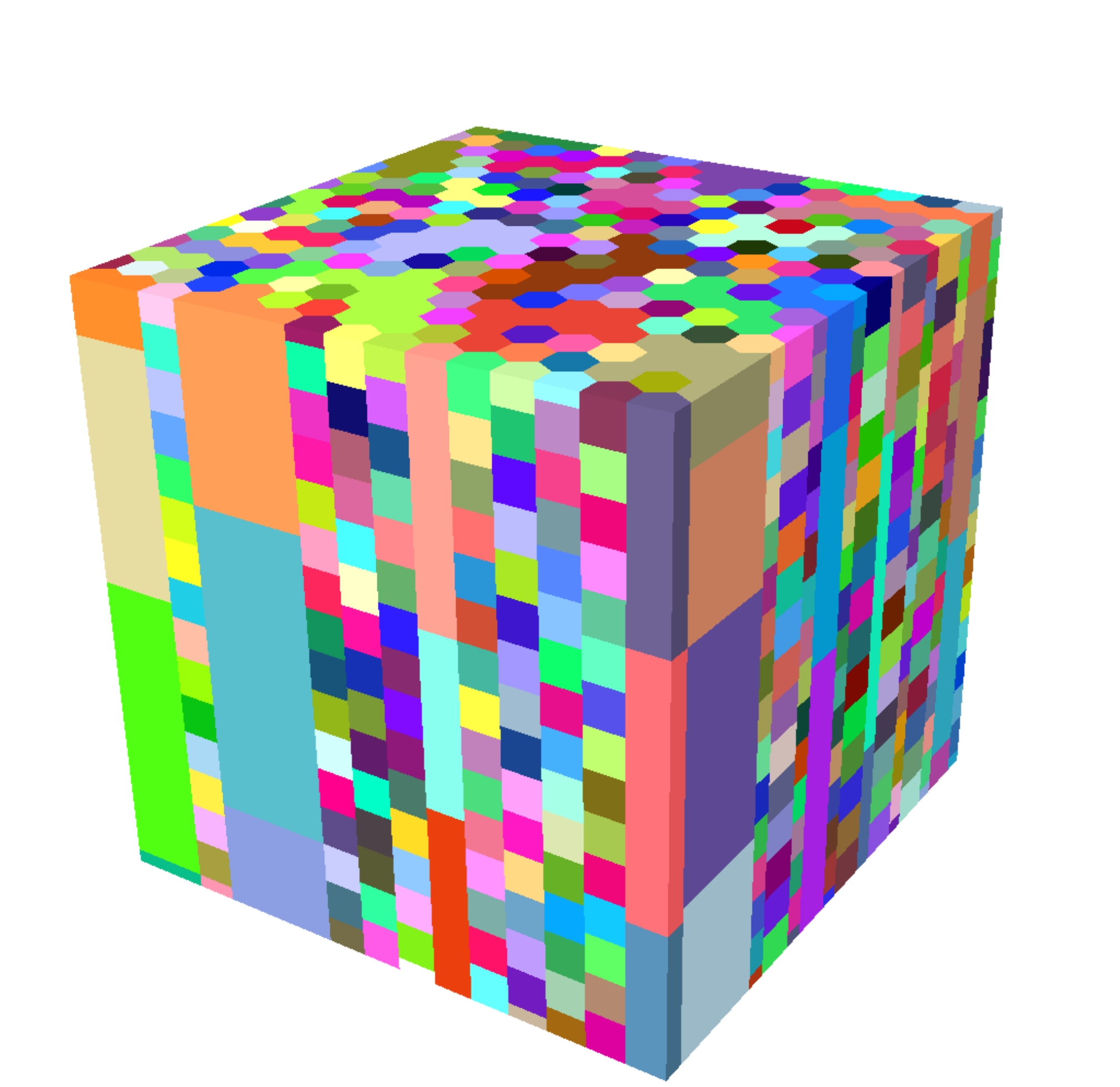}\label{fig:microstruct3D}}
	\caption{Microstructure generation mimics the material processing. Following the generation of the underlying mesh \subref{fig:base_mesh}, a two-dimensional microstructure is created. The two-dimensional parent ferrite microstructure is generated using mapped Voronoi tessellation \subref{fig:ferrite2D}. Austenite regions are then randomly transformed out along grain boundaries, until the desired volume ratio between the two phases is obtained \subref{fig:microstruct2D}. The two-dimensional microstructure is then extruded to form a three-dimensional microstructure \subref{fig:microstruct3D}. }
	\label{fig:MicrostructGen}
\end{figure}

Microstructure generation mimics the material processing, as illustrated in Figure~\ref{fig:MicrostructGen}. Following the generation of the underlying mesh, a two-dimensional microstructure is created. The two-dimensional parent ferrite microstructure is generated using mapped Voronoi tessellation. Austenite regions are then randomly transformed out along grain boundaries, until the desired volume ratio between the two phases is obtained. The two-dimensional microstructure is then extruded to form a three-dimensional microstructure. Each resulting columnar region is divided into grains based on a grain size distribution. All the grains in a column are assigned to the same phase as that of the parent two-dimensional region. Grain orientations are assigned by randomly sampling the experimentally-determined orientation distribution function for each phase.

\begin{figure}[h]
	\centering
	\subfigure[Grain]{\includegraphics[width=0.49\linewidth]{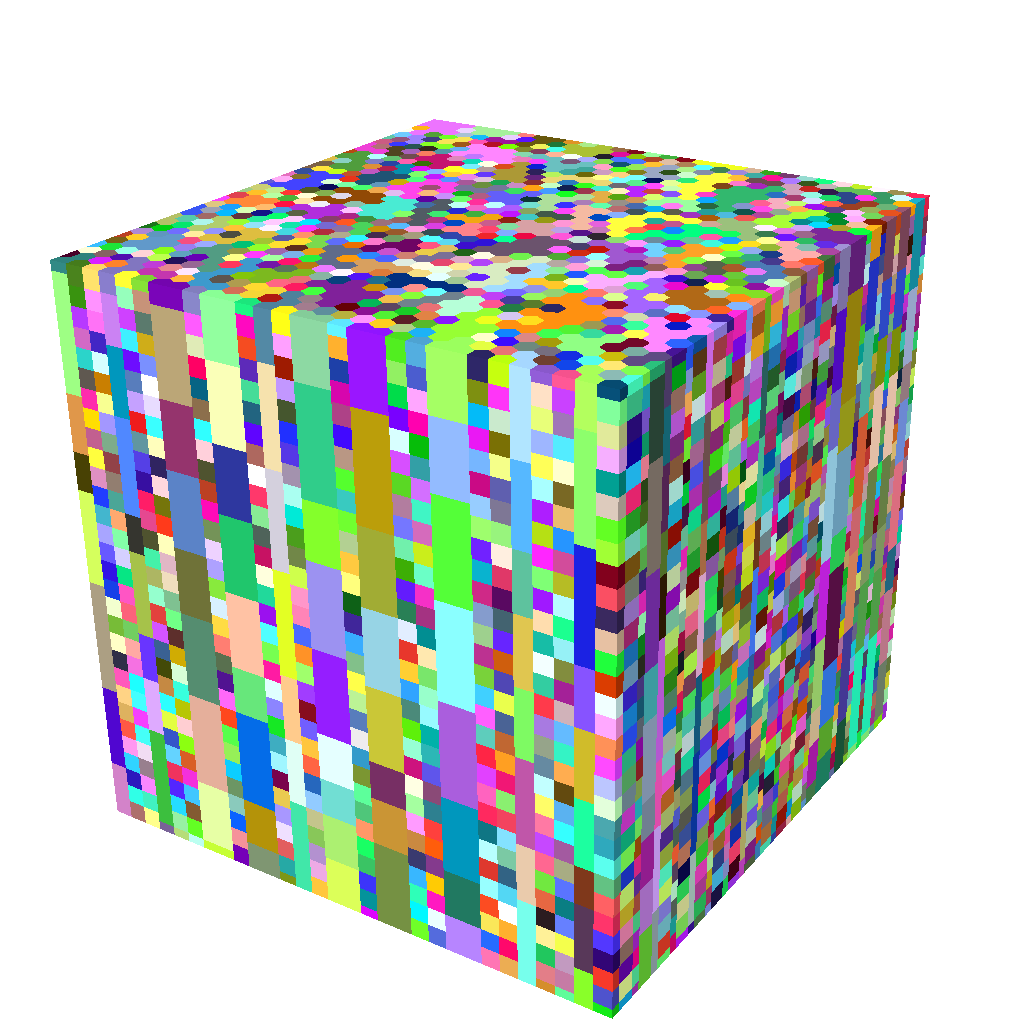}\label{fig:extuded-grain}}
	\subfigure[Phase]{\includegraphics[width=0.49\linewidth]{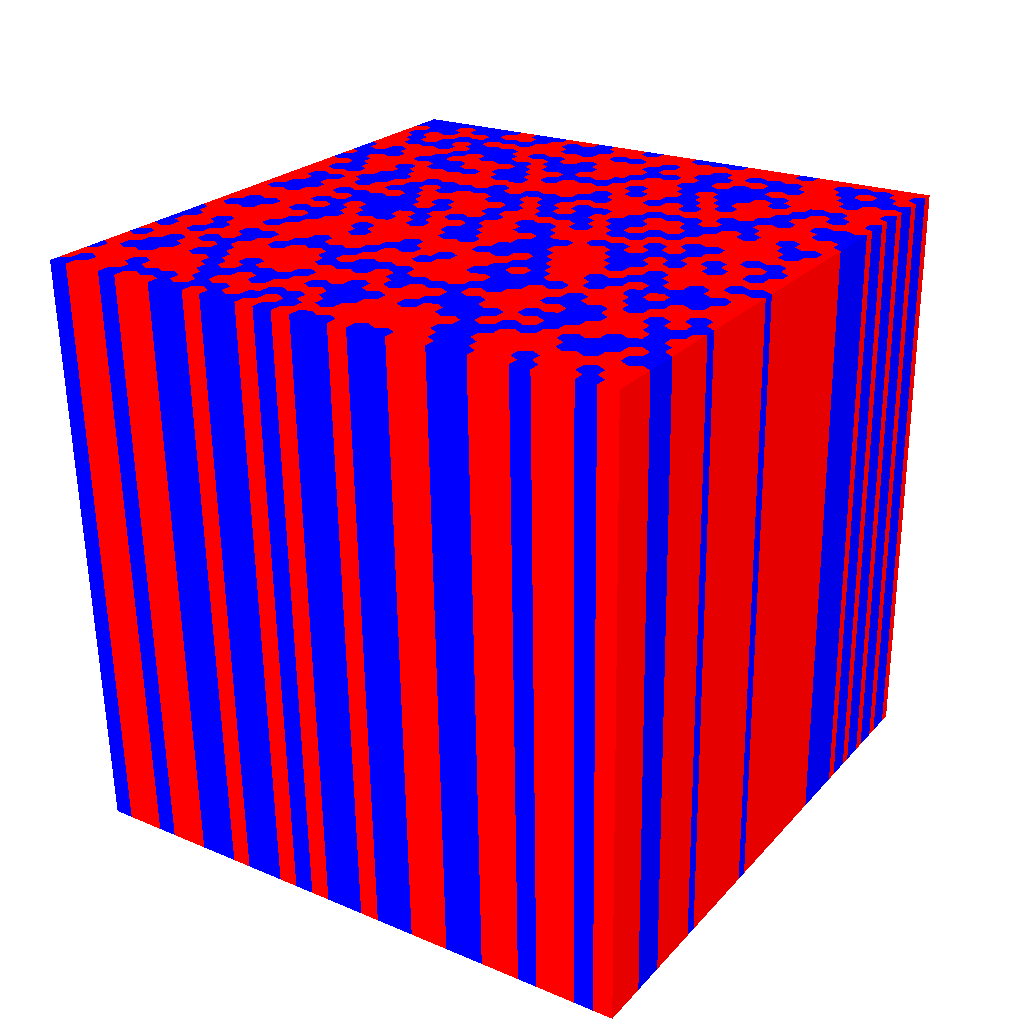}\label{fig:extruded-phase}}
	\caption{Rolled microstructure.}
	\label{fig:mesh-extruded}
\end{figure}

\begin{figure}[h]
	\centering
	\subfigure[Grain]{\includegraphics[width=0.49\linewidth]{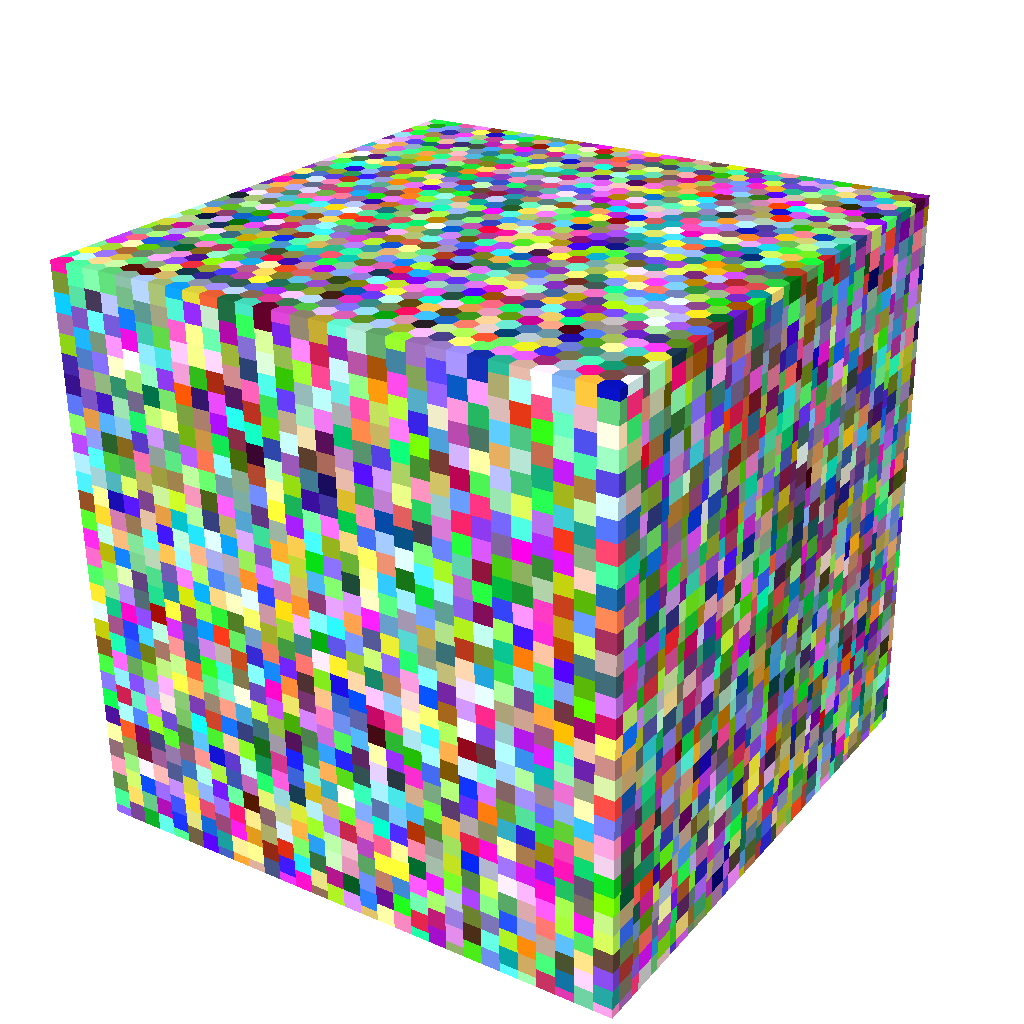}\label{fig:equiaxed-columnar-grain}}
	\subfigure[Phase]{\includegraphics[width=0.49\linewidth]{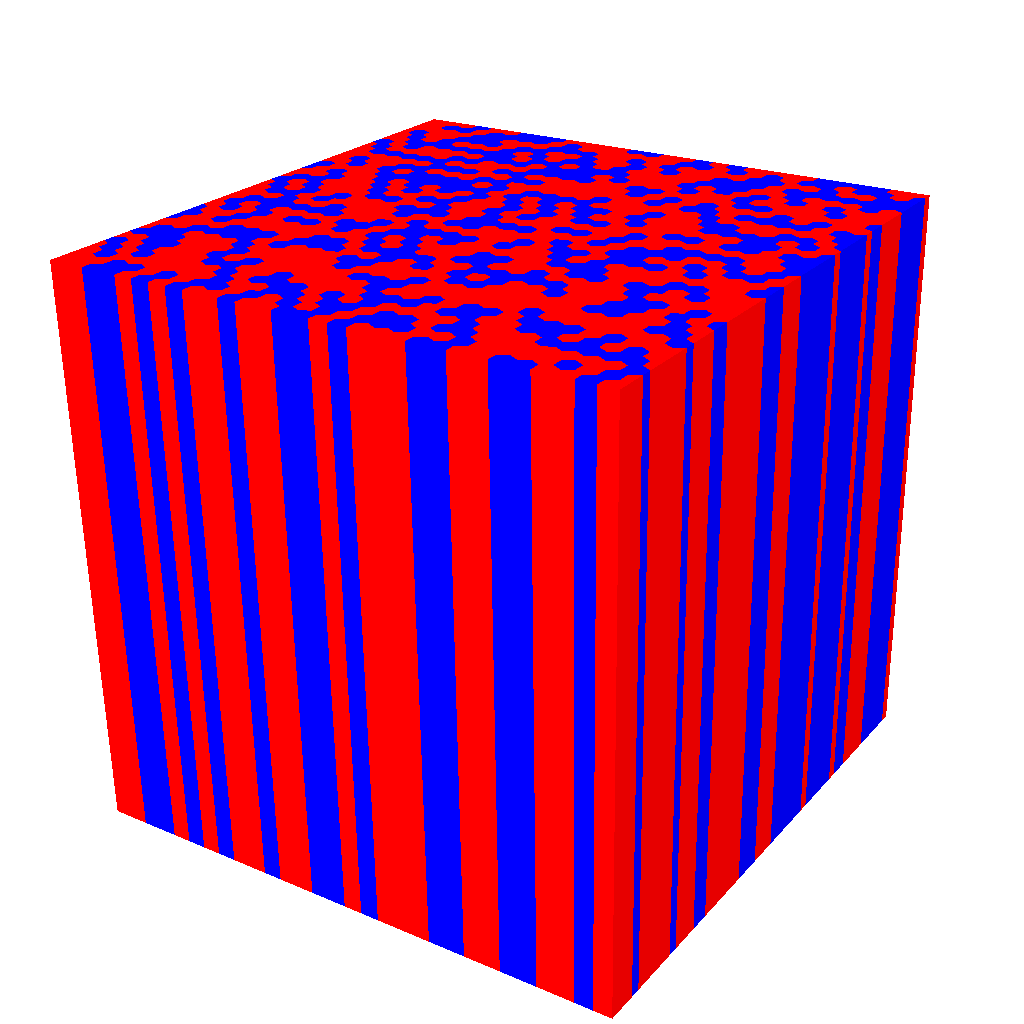}\label{fig:equiaxed-columnar-phase}}
	\caption{Equiaxed hexagonal grain, columnar phase microstructure.}
	\label{fig:mesh-equiaxed-columnar}
\end{figure}

\begin{figure}[h]
	\centering
	\subfigure[Grain]{\includegraphics[width=0.49\linewidth]{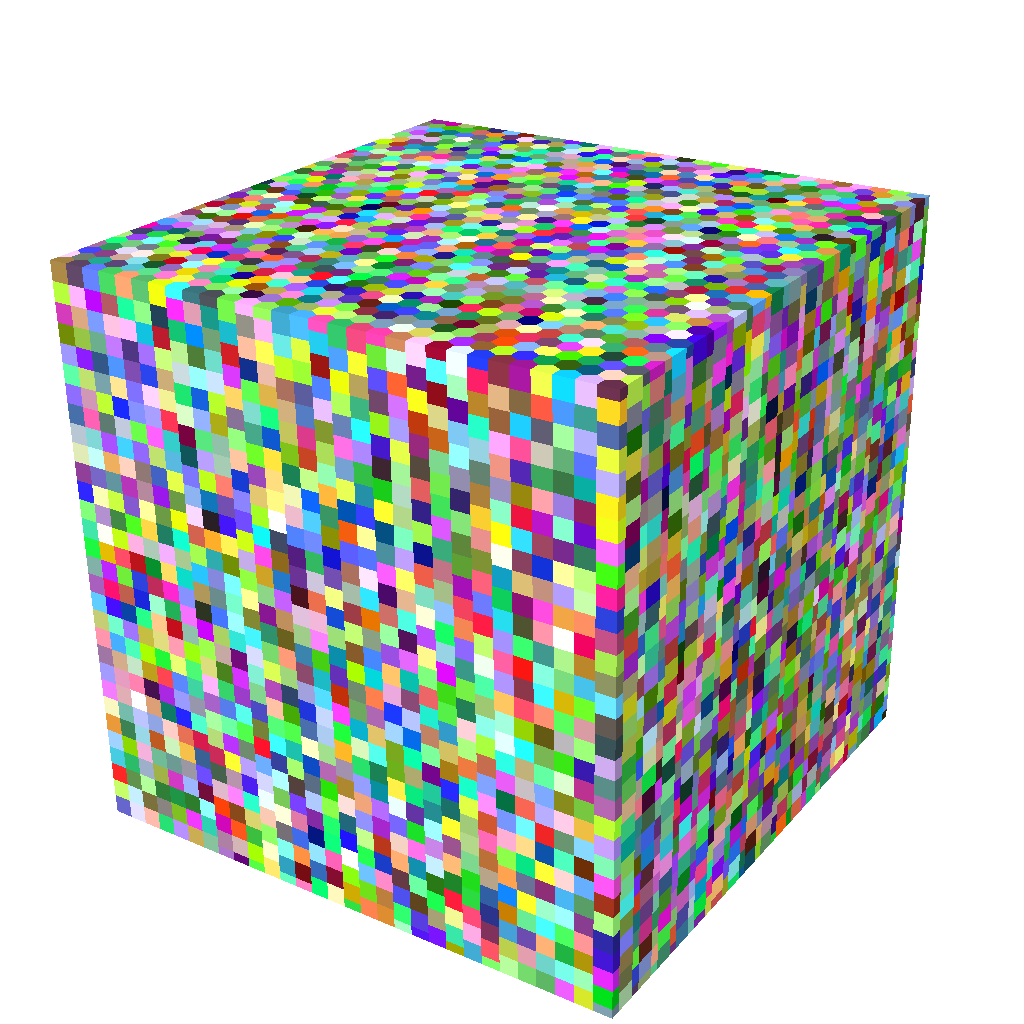}\label{fig:equiaxed-hex-grain}}
	\subfigure[Phase]{\includegraphics[width=0.49\linewidth]{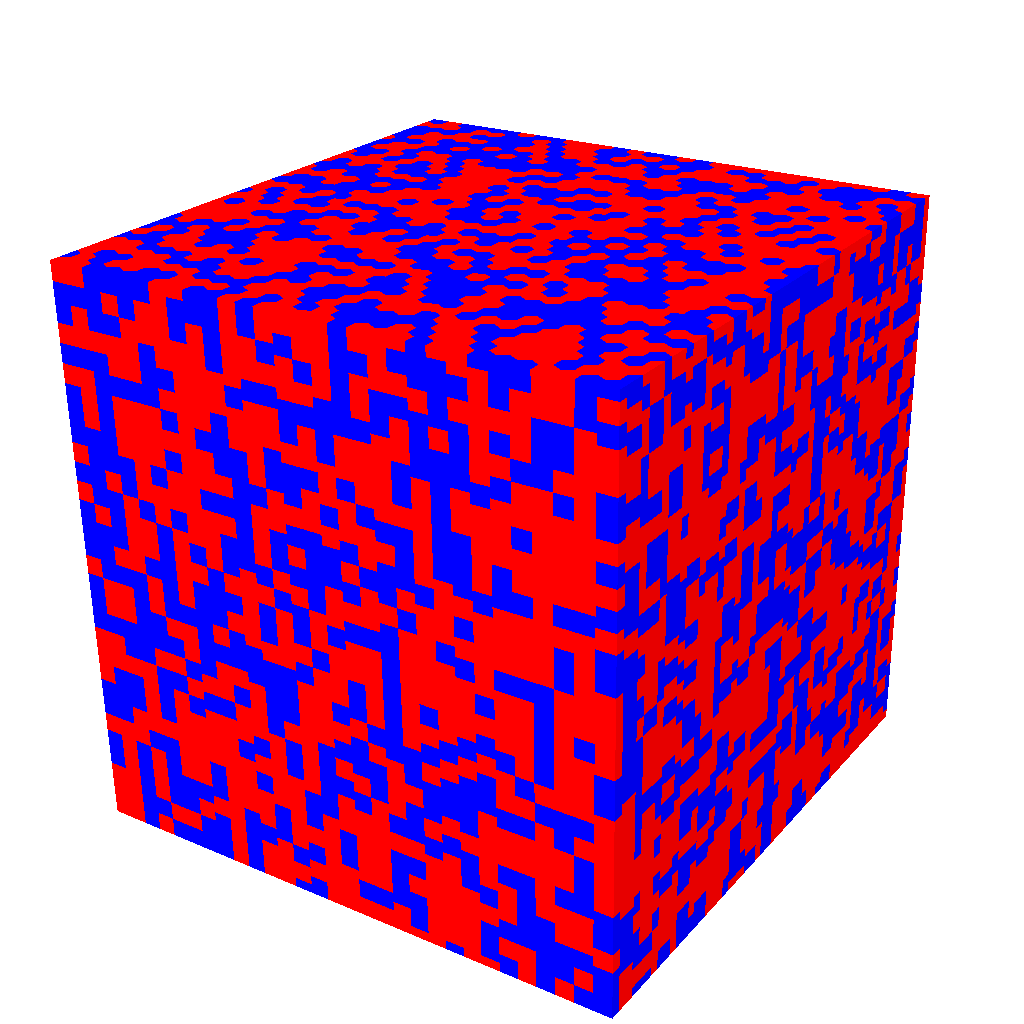}\label{fig:equiaxed-hex-phase}}
	\caption{Equiaxed hexagonal grain, random phase microstructure.}
	\label{fig:mesh-equiaxed-hex}
\end{figure}

\begin{figure}[h]
	\centering
	\subfigure[Grain]{\includegraphics[width=0.49\linewidth]{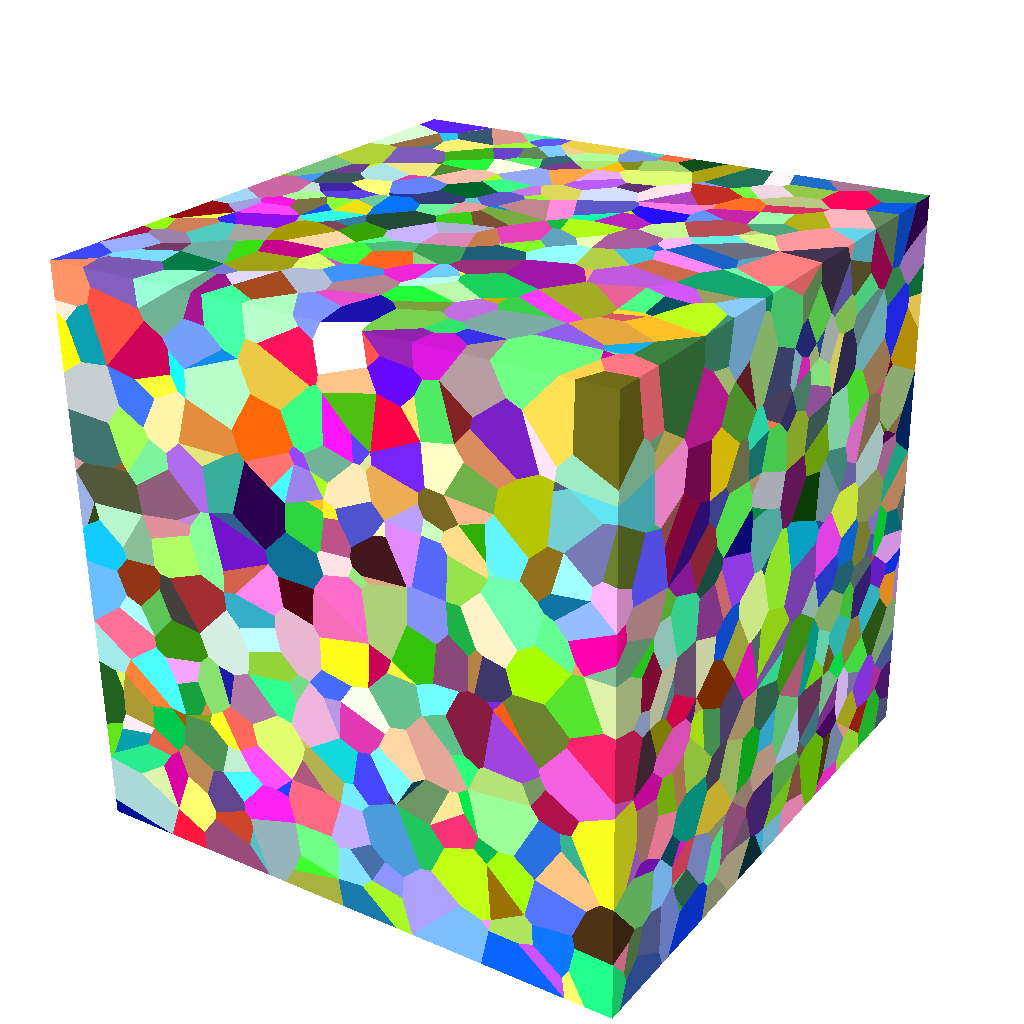}\label{fig:Voronoi-grain}}
	\subfigure[Phase]{\includegraphics[width=0.49\linewidth]{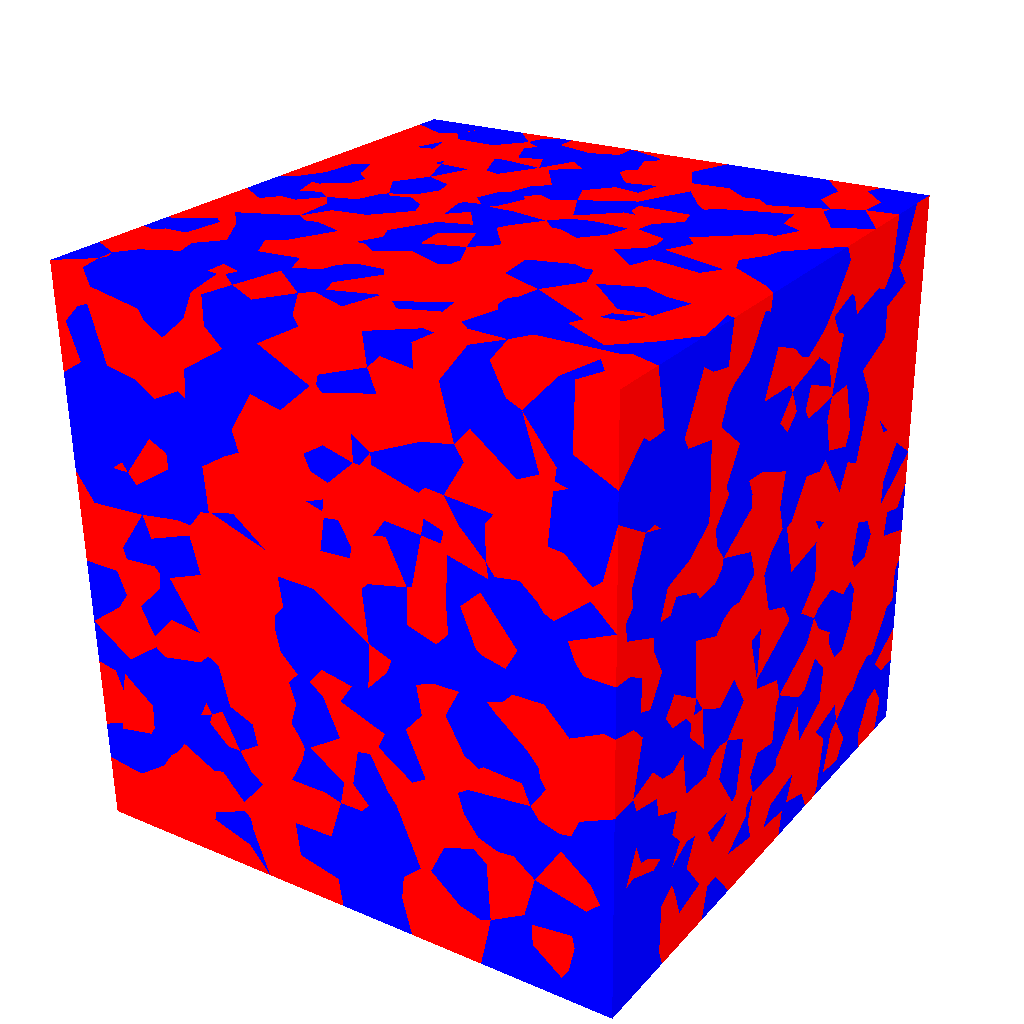}\label{fig:Voronoi-phase}}
	\caption{Equiaxed Voronoi microstructure.}
	\label{fig:mesh-Voronoi}
\end{figure}

Four microstructures types were considered. The rolled microstructure (Figure~\ref{fig:mesh-extruded}), which consists of a columnar phase structure, with large, irregularly shaped ferrite grains, most closely resembles that of LDX-2101. The equiaxed hexagonal grain, columnar phase microstructure (Figure~\ref{fig:mesh-equiaxed-columnar}) also has a columnar phase structure but is comprised of hexagonal prismatic grains. The equiaxed hexagonal grain, random phase microstructure (Figure~\ref{fig:mesh-equiaxed-hex}) lacks a columnar phase structure and is comprised of hexagonal prismatic grains. All three of these microstructures were created from different instantiations of the same underlying mesh. Lastly, an equiaxed Voronoi microstructure (Figure~\ref{fig:mesh-Voronoi}) was created with Voronoi tessellation using Neper~\cite{Quey11a}.

The mechanical responses of the four types of virtual polycrystals  were simulated using the finite element code, \fepx.   A summary of the formulation basis of \fepx\, is provided in the following subsection.  The identification of the material properties required of the formulation were discussed in Section~\ref{sec:parameters}.  The values used here in the examination of microstructure sensitivity are those identified in  Section~\ref{sec:parameters} and discussed in Section~\ref{sec:discussion}.

\clearpage

\subsection{Finite element formulation}\label{sec:fem_formulation}
The motion of a polycrystal is determined by solving a set of field equations consisting of equations of equilibrium and kinematics, equations for the constitutive behavior, and boundary conditions. 
In this work, we use the finite element code, \fepx, to solve this
set of equations specifically for the mechanical response polycrystalline aggregates.  The summary provided here of the \fepx\, formulation is taken heavily from its documentation~\cite{Dawson14a}.

Equilibrium is enforced by requiring a global weighted residual to vanish:
\begin{equation}
R_u = \int_{\mathcal{B}} \gtnsr{\psi} \cdot \left( \divop {\cauchy}^T + {\vctr{\iota}}\right) \dee \mathcal{B} = 0
\end{equation}
The residual is manipulated in the customary manner (integration by parts and application of the divergence theorem) to obtain the weak form:
\begin{equation}
R_u = 
- \int_{\mathcal{B}} \; \trace  \left( {\dcauchy}^\transp  \,\gradop \vctr{\psi} \right) \dee {\mathcal{B}} 
+  \int_{\mathcal{B}} \pi \, \divop \vctr{\psi} \dee {\mathcal{B}}
+ \int_{\partial {\mathcal{B}}} \vctr{t} \cdot \vctr{\psi} \dee\Gamma +
\int_{{\mathcal{B}}} \vctr{\iota} \cdot \vctr{\psi} \dee{\mathcal{B}}
\end{equation}
Kinematics are introduced through equations that relate the velocity field to the deformation rate and spin via the velocity gradient, as outlined in Section~\ref{sec:model_formulation}.  The velocity field is represented with the finite element interpolation functions presented later in the appendix.

The constitutive relations enter to relate the stress to the motion, in particular the deformation rate and spin.  
 Equations~\ref{eq:trD} to \ref{eq:schmidtensor} are now merged into a single equation that relates the Cauchy stress to the
 total deformation rate.   First,  the spatial time-rate change of the elastic strain is approximated with a finite difference expression:
\begin{equation}
\matlatepsdot = \deeteei \biggl( \matlateps - \matlatepsold \biggr)
\label{eq:euler_approx_strain_rate}
\end{equation}
where $\matlateps$ is the elastic strain at the end of the time step and $\matlatepsold$ is the elastic strain at the beginning of the time step.   The difference approximation is employed in an implicit algorithm, wherein the equations are solved at the time corresponding to the end of the time step.  This time corresponds to the current configuration.  
Writing the time rate change of the strain in terms of strains at two times facilitates substitution of Hooke's law -- namely at the end of the time step.   The elastic strain at the beginning of the time step is known from the solution for the preceding time step.
For the volumetric part of the motion this gives:
\begin{equation}
-\pi  =  \frac{\kappa\Delta t }{ \beta } \trace\matdefrate +  \frac{\kappa} {\beta} \trace\matlatepsold
\label{eq:discret_volumetric_ep-law}
\end{equation}
For the deviatoric part, we obtain:
\begin{equation}
\matddefrate = \deeteei \matdlateps + \matlatdefrate + \matpxspinhat \matdlateps - \deeteei \matdlatepsold
\label{eq:discret_deviatoric_ep-law}
\end{equation}
where $\matpxspinhat$ is the matrix form of $\pxspinhat$:
\begin{equation}
\matpxspinhat = 
\left[
\begin{array}{c c c c c}
0 &  0 & -2{\hat w_{12}^p} & -{\hat w_{13}^p} &  {\hat w_{23}^p} \\ 
0 & 0 & 0 & {\sqrt{3}} {\hat w_{13}^p} & {\sqrt{3}} {\hat w_{23}^p}  \\ 
2{\hat w_{12}^p}& 0 & 0 & -{\hat w_{23}^p}  &  -{\hat w_{13}^p}  \\ 
{\hat w_{13}^p}  & -{\sqrt{3}}{\hat w_{13}^p}  &  {\hat w_{23}^p} & 0  &  -{\hat W_{12}^p} \\  
-{\hat w_{23}^p}  & -{\sqrt{3}}{\hat w_{23}^p}  &  {\hat w_{13}^p} & {\hat w_{12}^p}   &  0 
\end{array}
\right] 
\end{equation}
The equations for plastic slip define the plastic deformation rate in terms of the Kirchhoff stress:
\begin{equation}
\matlatdefrate = \matxplasticity \matdkirch
\end{equation}
\begin{equation}
\matxplasticity
= 
\sumss \left( \frac{f(\rss, g)}{\rss} \right) 
\matsymschmid  \matsymschmid^\transp
\end{equation}
When combined with Hooke's law for the elastic response we obtain:
are substituted to render an equation that gives the deviatoric Cauchy stress in terms of the total deviatoric deformation rate and 
a matrix, $\mathhh$, that accounts for the spin and the elastic strain at the beginning of the time step: 
\begin{equation}
\matdcauchy  =  \matxep \bigg( \matddefrate- \mathhh \bigg)
\label{eq:discrete_devCauchy}
\end{equation}
where:
\begin{equation}
\matxep^{\invrs} = \frac{\beta}{\Delta t} {\matxdelasticity}^{\invrs} +\beta \matxplasticity 
\label{eq:discrete_ep_stiffness}
\end{equation}
\begin{equation}
\mathhh = \matpxspinhat  \matdlateps - \deeteei \matdlatepsold
\label{eq:discrete_spin_correction}
\end{equation}
Equations~\ref{eq:discrete_devCauchy}, \ref{eq:discrete_ep_stiffness} and \ref{eq:discrete_spin_correction} are substituted in the weak form of equilibrium to write the stress in terms of the deformation rate.

\fepx\, employs a standard isoparametric mapping framework for discretizing the problem domain and for representing the solution variables.  
The mapping of the coordinates of points provided by the elemental interpolation functions, $\matcapN$, and the coordinates of the nodal points, $\left\{ X \right\}$:
\begin{equation}
\left\{ x \right\}  = \matcapN  \left\{ X \right\}
\label{eq:coord_mapping}
\end{equation}
where $(\xi, \eta, \zeta)$ are local coordinates within an element.  
The same mapping functions are used for the solution (trial) functions which, together with the nodal point values of the
velocity, $\matvelnp$, specify the velocity field over the elemental domains:
\begin{equation}
\matvel  = \matcapN  \matvelnp
\label{eq:trial_functions}
\end{equation}
The deformation rate is computed from the spatial derivatives (derivatives with respect to $\vctr{x}$) of the mapping functions and the nodal velocities as:
\begin{equation}
\matdefrate = \matcapB \matvelnp
\label{eq:trial_function_derivatives}
\end{equation}
$\matcapB$ is computed using the derivatives of $\matcapN$ with respect to local coordinates, $(\xi, \eta, \zeta)$, together with the Jacobian of the mapping specified by Equation~\ref{eq:coord_mapping}, following standard finite element procedures for isoparametric elements.

\begin{figure}[ht]
\begin{center}
\includegraphics*[width=10cm]{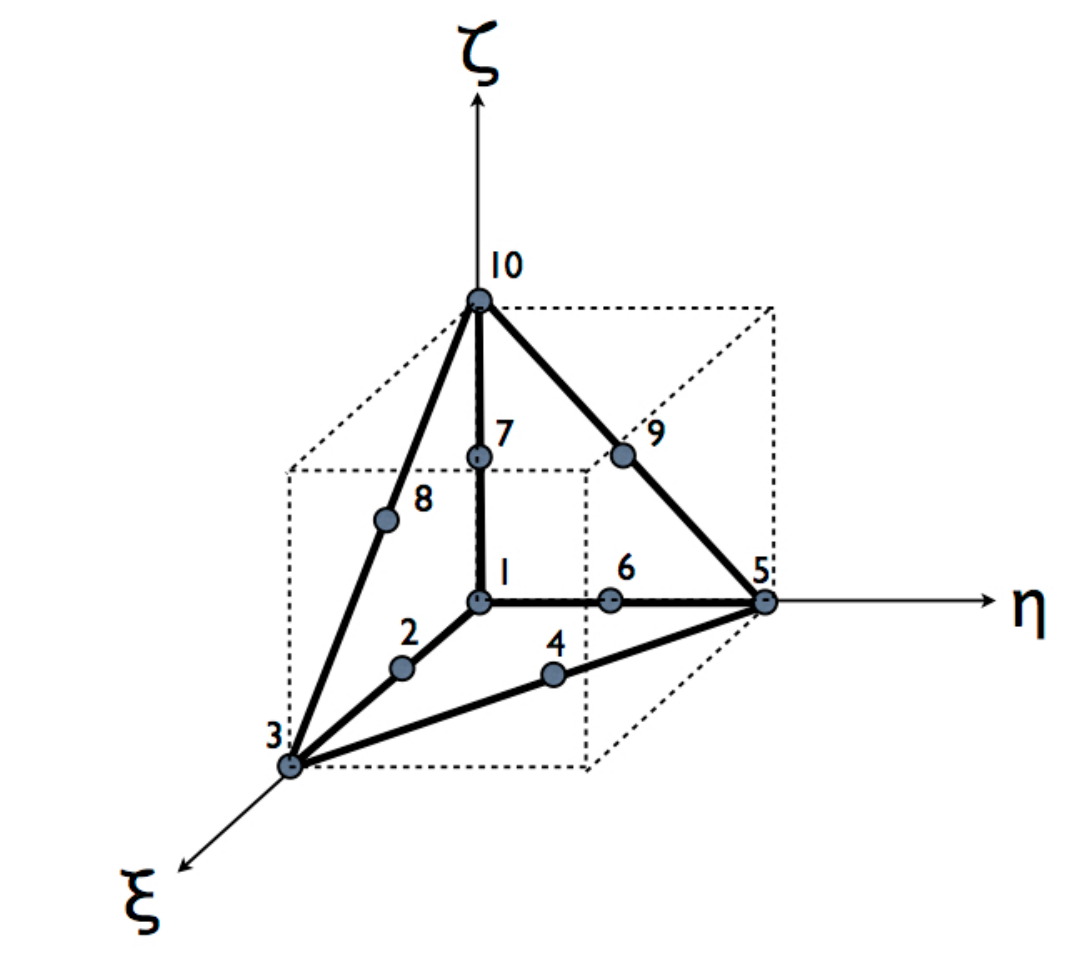}
\caption{10-node tetrahedral element with quadratic interpolation of the velocity, shown in the parent configuration and bounded by a unit cube.}
\label{fig:tet_element}
\end{center}
\end{figure}

\fepx\, relies principally on a 10-node, tetrahedral, serendipity element, as shown in Figure~\ref{fig:tet_element}.  This $C^0$ element provides pure  quadratic interpolation of the velocity field.  
\fepx\, employs a Galerkin methodology for constructing a weighted residual.  The weight functions therefore use the same interpolation functions as used for the coordinate map and the trial functions:
\begin{equation}
\Big\{ \psi \Big\} =  \matcapN  \Big\{ {\it \Psi } \Big\}
\label{eq:weight_functions}
\end{equation}

Introduction of the trial and weight functions gives a  residual vector for the discretized weak form for each element:  
\begin{equation}
\matresiduale
=
\bigg[ \matstiffd + \matstiffv \bigg] \matvelnp
-
\Big\{ \mathsf{f}^{\it ele}_a \Big\}
-
\Big\{ \mathsf{f}^{\it ele}_d \Big\}
-
\Big\{ \mathsf{f}^{\it ele}_v \Big\}
\end{equation}
where
\begin{equation}
\matstiffd 
= 
\int_{\mathcal{B}} \matcapB^\transp  \matcapX^\transp \matxep \matcapX \matcapB \dee {{\mathcal{B}}}
\label{eq:k_d}
\end{equation}
\begin{equation}
\matstiffv = 
\int_{\mathcal{B}} \betaoverkappa \matcapB^\transp \matcapX^\transp \matdelta \matdelta^\transp \matcapX \matcapB \dee{\mathcal{B}}
\end{equation}
\begin{equation}
\Big\{ \mathsf{f}^{\it ele}_a \Big\} 
= 
\int_{\partial{\mathcal{B}}}\matcapN^\transp \matbodyforce \dee {\mathcal{B}}
\end{equation}
\begin{equation}
\Big\{ \mathsf{f}^{\it ele}_v \Big\} 
= \int_{\mathcal{B}} \matcapB^\transp \matcapX^\transp \betaoverkappa \matdelta^\transp \matlatepsold \dee{\mathcal{B}} 
\end{equation}
\begin{equation}
\Big\{ \mathsf{f}^{\it ele}_d \Big\} 
=  \int_{\mathcal{B}} \matcapB^\transp \matcapX^\transp \matxep \mathhh \dee{\mathcal{B}} 
\label{eq:f_d}
\end{equation}
The $\matdelta$ and $\matcapX$ matrices facilitate performing trace and inner product operations.  For details, see \cite{Dawson14a}.
The integrals appearing in Equations~\ref{eq:k_d}-\ref{eq:f_d}  are evaluated by numerical quadrature.

An iterative procedure is used in \fepx\, to advance the coupled solutions for the motion and state over a specified history of loading as described in Section~\ref{sec:parameters}.

\subsection{Mechanical responses of virtual polycrystals}\label{sec:vp_instantiations}

\begin{figure}[h]
	\centering
	\subfigure[Full.]{\includegraphics{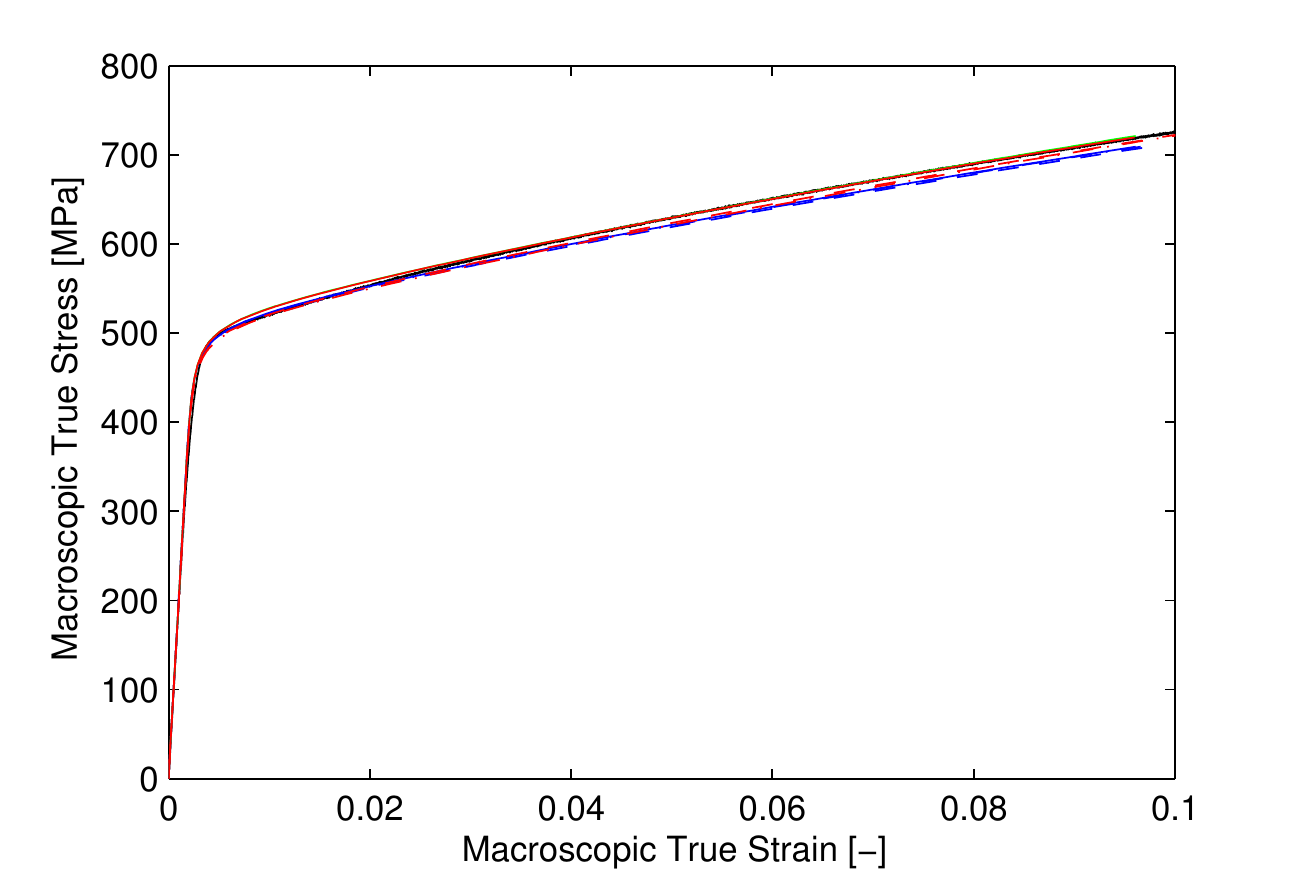}\label{fig:sigeps_microstruct_ldx_full}}
	\subfigure[Zoom.]{\includegraphics{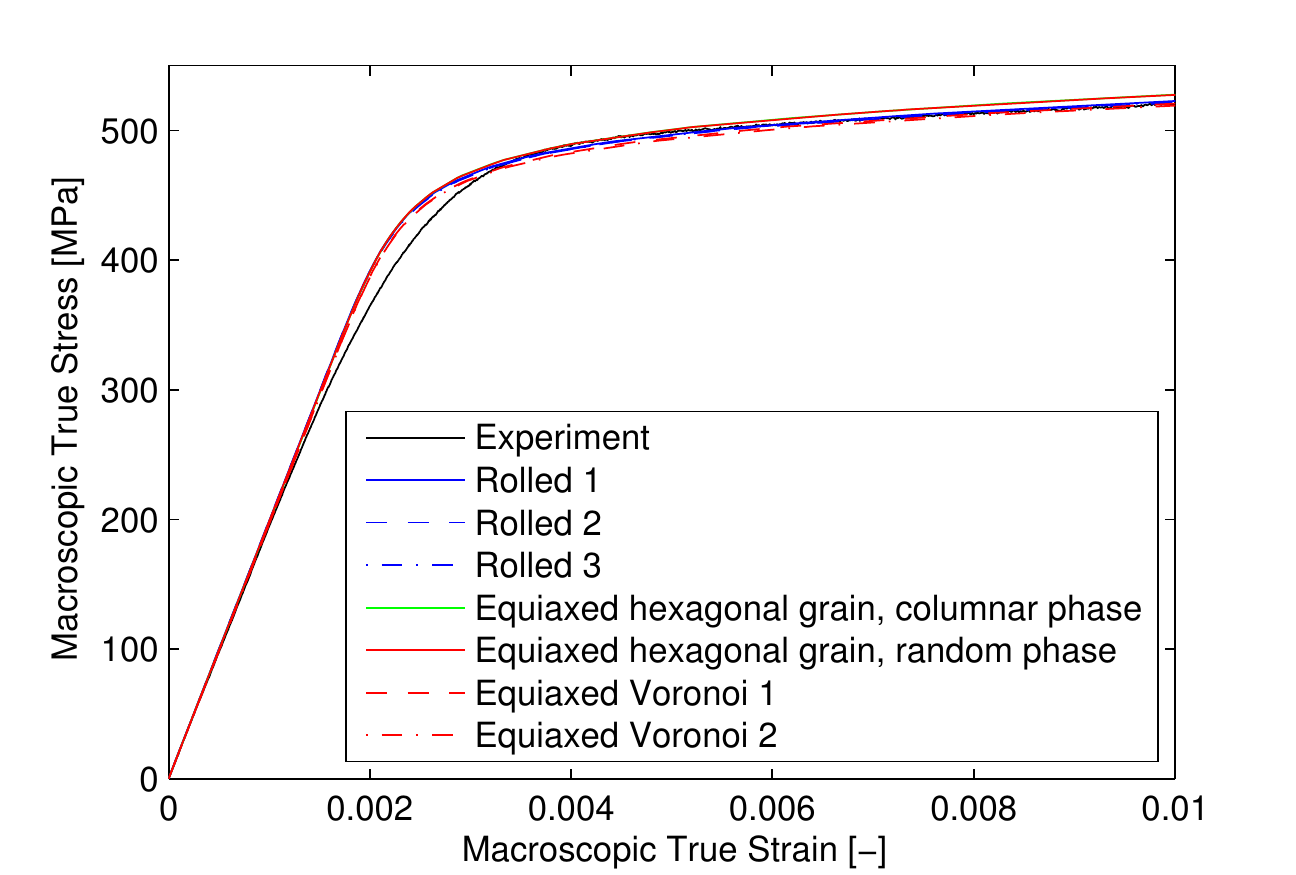}\label{fig:sigeps_microstruct_ldx_zoom}}
	\caption{Macroscopic stress-strain response for four types of microstructures with LDX-2101 material parameters.}
	\label{fig:sigeps_microstruct_ldx}
\end{figure}

\begin{figure}[h]
	\centering
	\subfigure[FCC \{200\}.]{\includegraphics[width=0.49\linewidth]{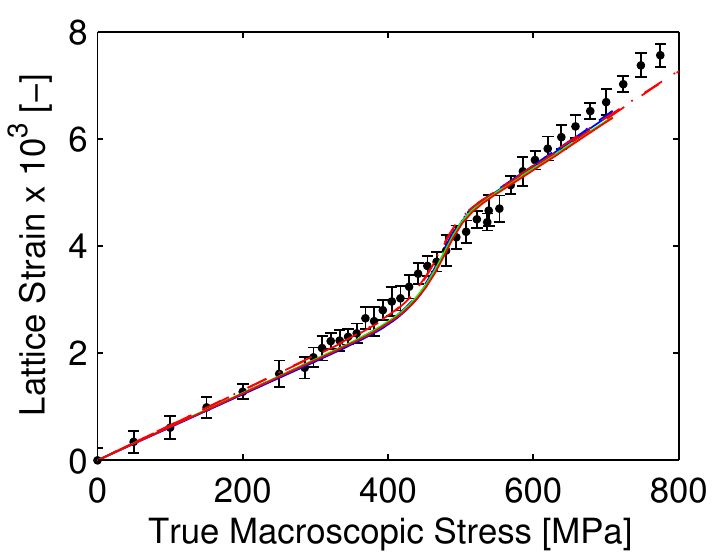}\label{fig:LSMicrostructFCC200}}
	\subfigure[BCC \{200\}.]{\includegraphics[width=0.49\linewidth]{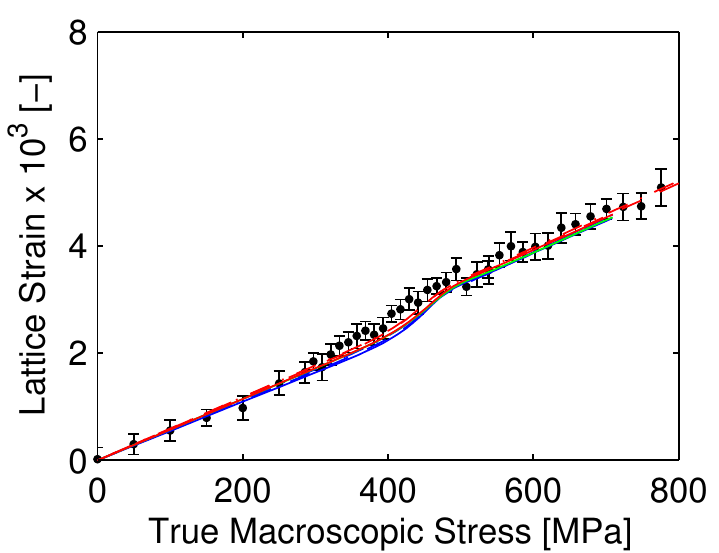}\label{fig:LSMicrostructBCC200}}
	\subfigure[FCC \{111\}.]{\includegraphics[width=0.49\linewidth]{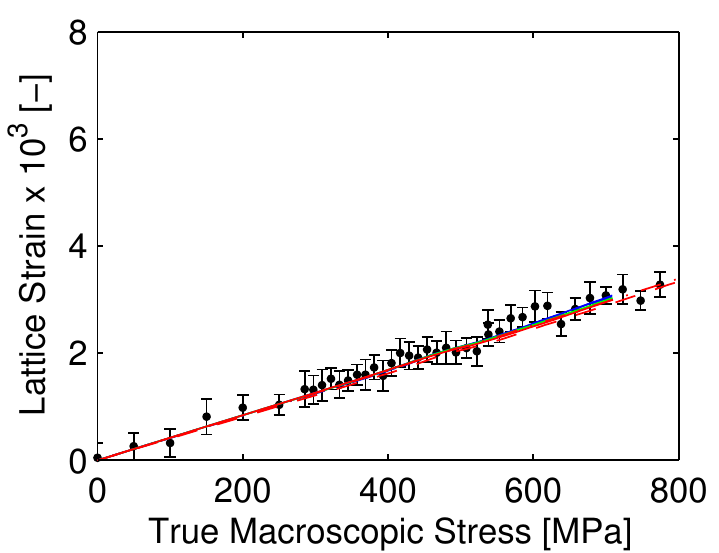}\label{fig:LSMicrostructFCC222}}
	\subfigure[BCC \{110\}.]{\includegraphics[width=0.49\linewidth]{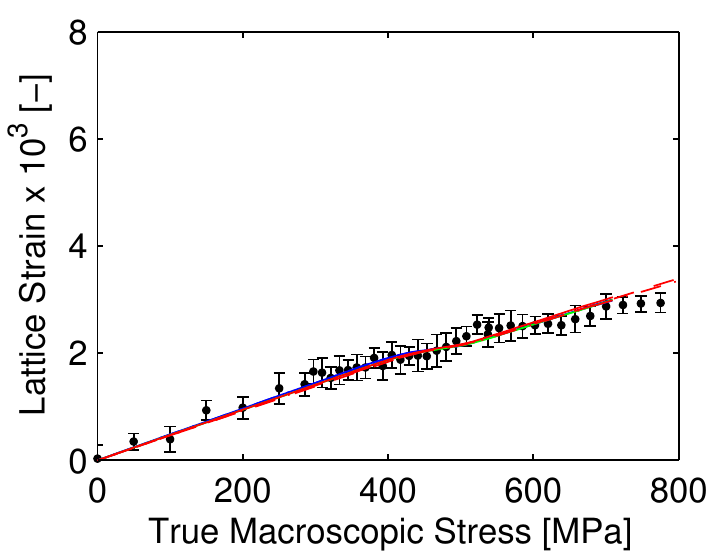}\label{fig:LSMicrostructBCC220}}
	\subfigure[FCC \{220\}.]{\includegraphics[width=0.49\linewidth]{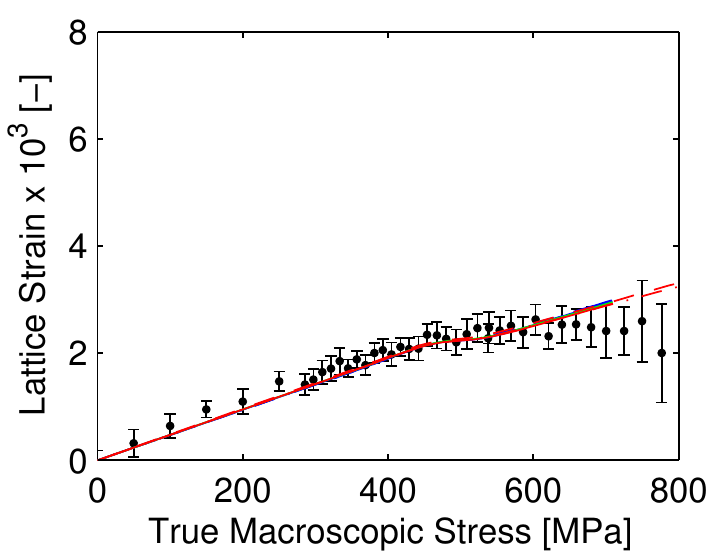}\label{fig:LSMicrostructFCC220}}
	\subfigure[BCC \{211\}.]{\includegraphics[width=0.49\linewidth]{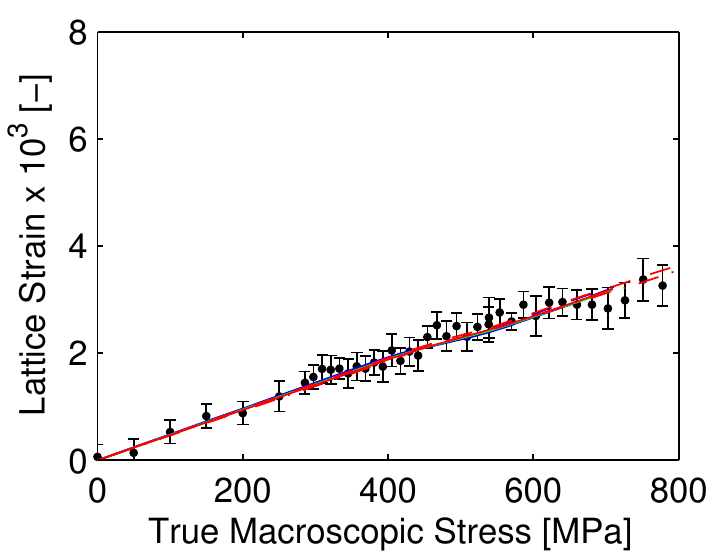}\label{fig:LSMicrostructBCC211}}
	\caption{Fiber-averaged axial lattice strain responses for four different types of microstructures with LDX-2101 material parameters. Refer to the legend in Figure~\ref{fig:sigeps_microstruct_ldx}.}
	\label{fig:LSMicrostruct}
\end{figure}

Between one and three instantiations of each microstructure were generated. Simulations were conducted for monotonic, uniaxial tension at a constant strain rate of $10^{-4}~\mathrm{s}^{-1}$, using material parameters for LDX-2101. The macroscopic stress-strain response (Figure~\ref{fig:sigeps_microstruct_ldx}) exhibits little microstructure dependence. Similarly, the fiber-averaged lattice strains are also relatively insensitive to microstructure (Figure~\ref{fig:LSMicrostruct}). The relative insensitivity of the mechanical behavior to material microstructure can be attributed to the fact that both phases have similar material properties. Therefore, in this case, the spatial arrangement of the two phases does not have a significant effect on bulk or fiber-averaged material behavior.

\begin{figure}[h]
	\centering
	\subfigure[Full.]{\includegraphics{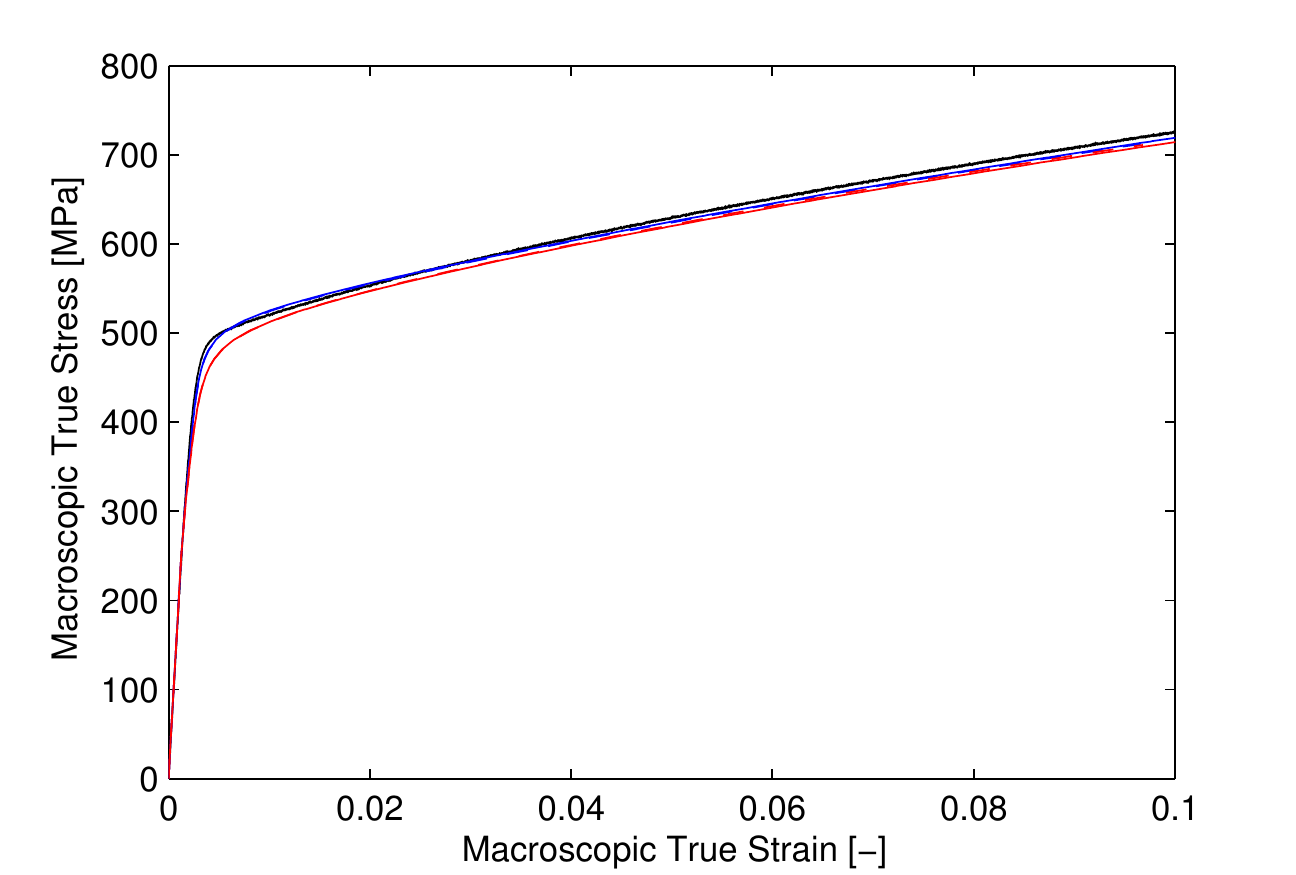}\label{fig:sigeps_microstruct_strength_full}}
	\subfigure[Zoom.]{\includegraphics{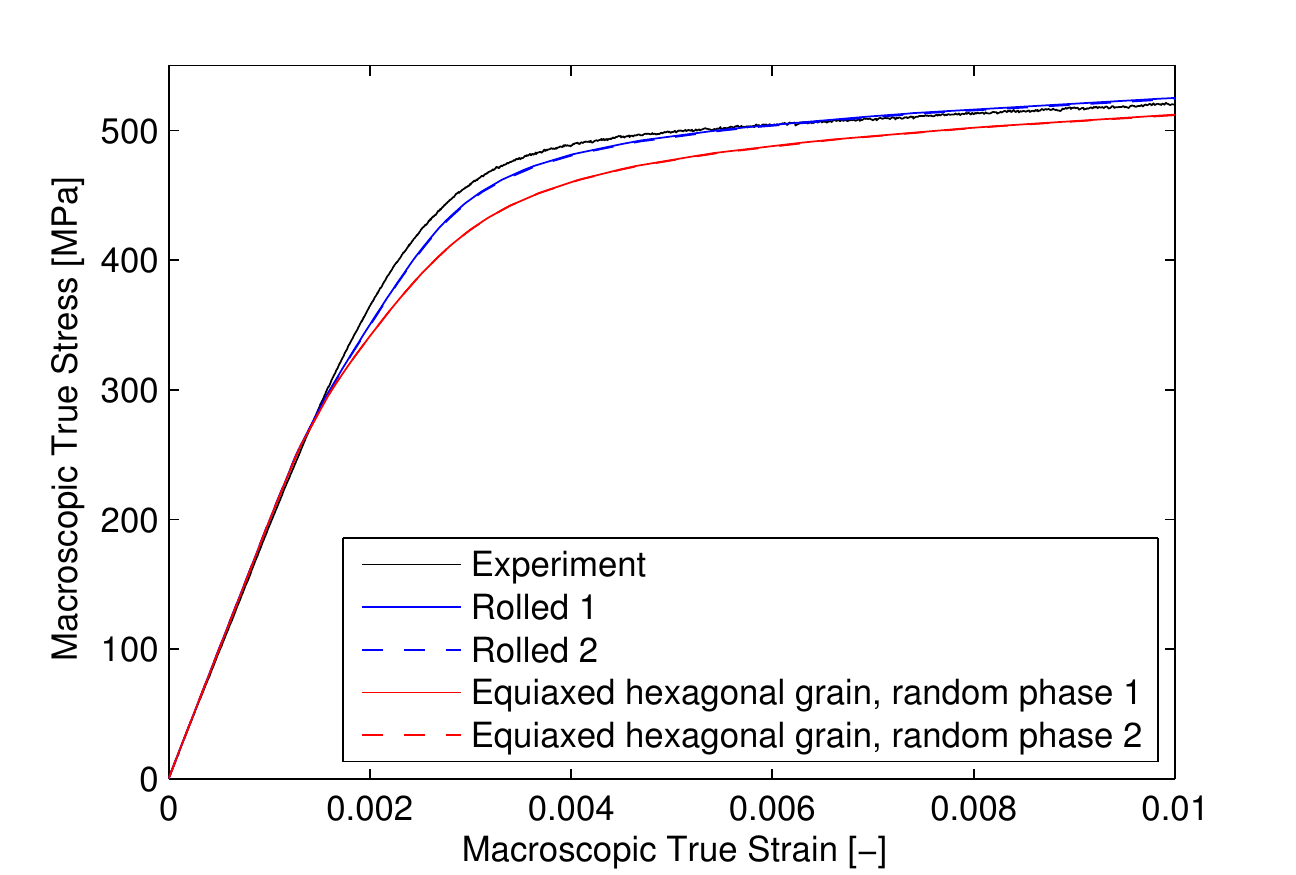}\label{fig:sigeps_microstruct_strength_zoom}}
	\caption{Macroscopic stress-strain response for four types of microstructures with the initial slip system strength ($g_0$) of austenite twice that of ferrite.}
	\label{fig:sigeps_microstruct_strength}
\end{figure}

\begin{figure}[h]
	\centering
	\subfigure[FCC \{200\}.]{\includegraphics[width=0.49\linewidth]{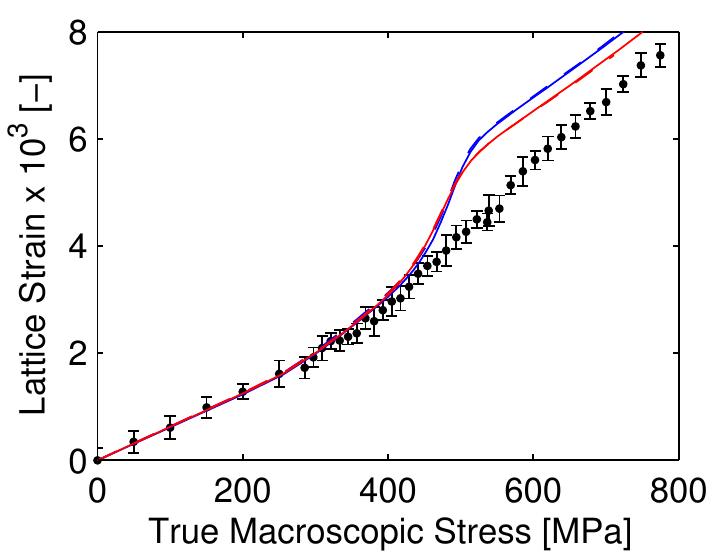}\label{fig:LSMicrostructStrengthFCC200}}
	\subfigure[BCC \{200\}.]{\includegraphics[width=0.49\linewidth]{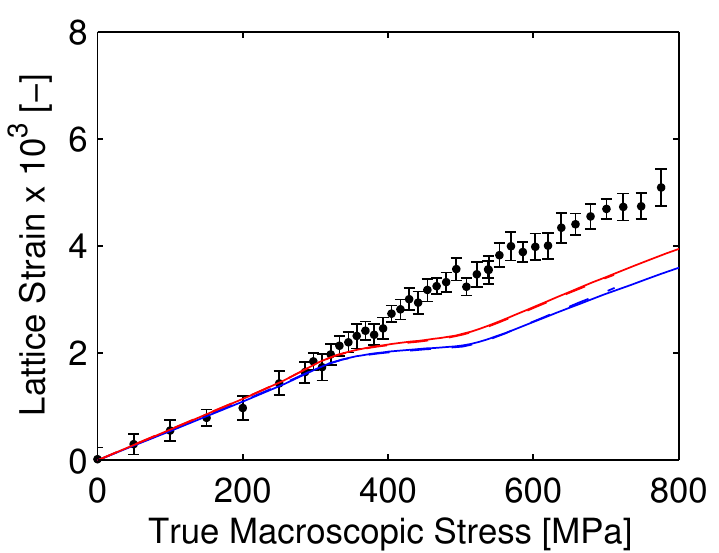}\label{fig:LSMicrostructStrengthBCC200}}
	\subfigure[FCC \{111\}.]{\includegraphics[width=0.49\linewidth]{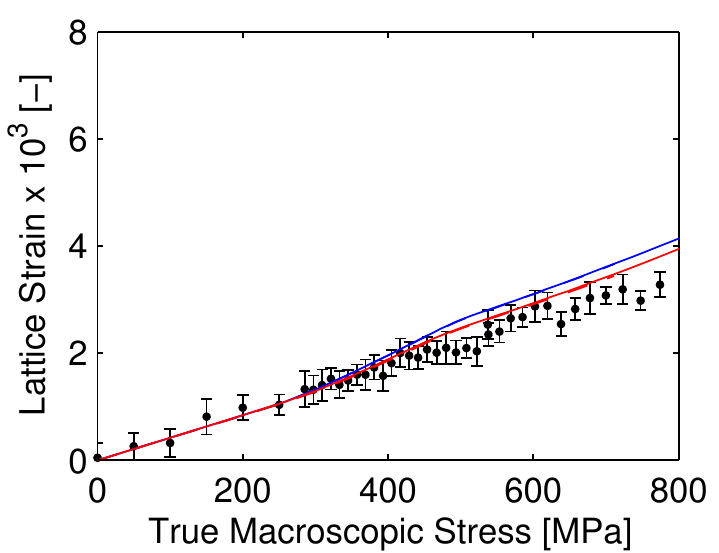}\label{fig:LSMicrostructStrengthFCC222}}
	\subfigure[BCC \{110\}.]{\includegraphics[width=0.49\linewidth]{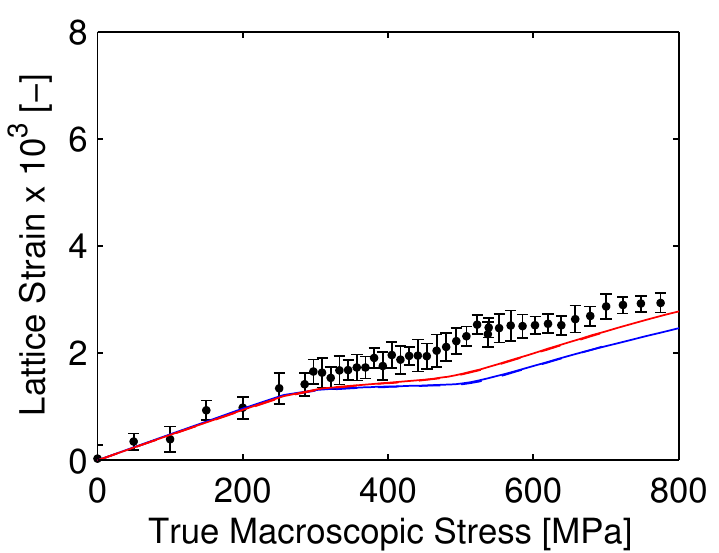}\label{fig:LSMicrostructStrengthBCC220}}
	\subfigure[FCC \{220\}.]{\includegraphics[width=0.49\linewidth]{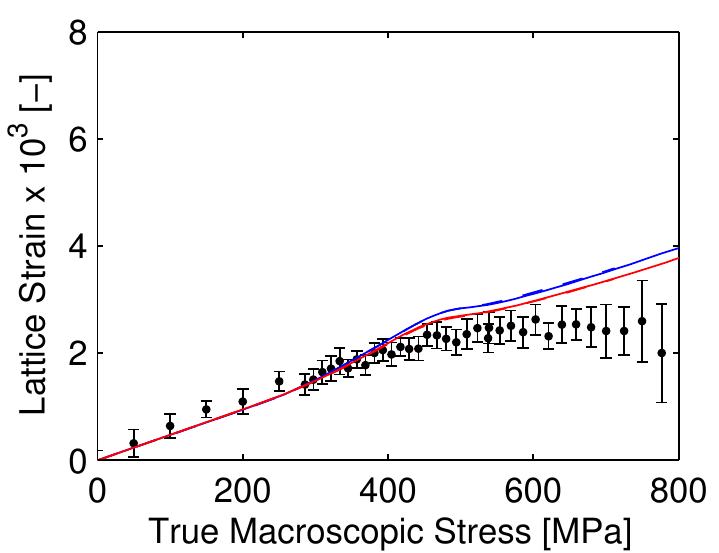}\label{fig:LSMicrostructStrengthFCC220}}
	\subfigure[BCC \{211\}.]{\includegraphics[width=0.49\linewidth]{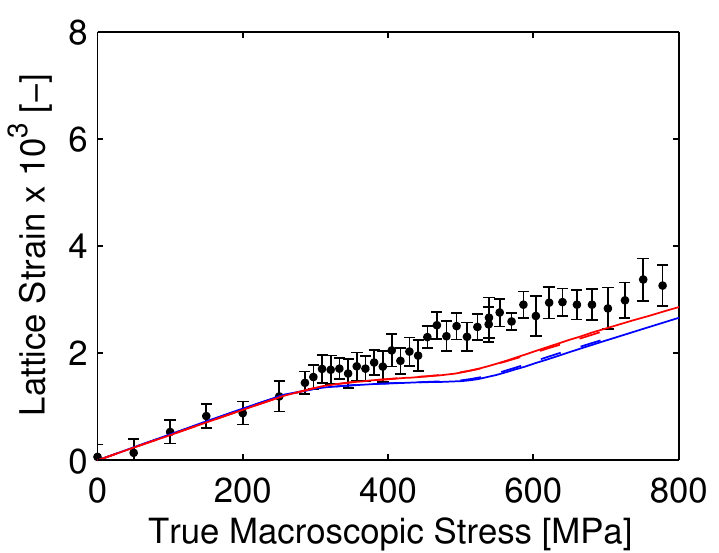}\label{fig:LSMicrostructStrengthBCC211}}
	\caption{Fiber-averaged axial lattice strain responses for four different types of microstructures with the initial slip system strength ($g_0$) of austenite twice that of ferrite. Refer to the legend in Figure~\ref{fig:sigeps_microstruct_strength}.}
	\label{fig:LSMicrostructStrength}
\end{figure}

To investigate whether microstructure has a greater effect on mechanical response when the phases have dissimilar properties, simulations were repeated with the initial slip system strength ($g_0$) of austenite twice that of ferrite. The volume-averaged initial slip system strength was the same as for LDX-2101. The major difference in macroscopic stress-strain response occurs during the elasto-plastic transition (Figure~\ref{fig:sigeps_microstruct_strength}). The yield strength, based on the 0.2\% offset method, is reduced by 20~MPa for the equiaxed hexagonal grain, random phase microstructure. The rolled microstructure is stronger because of the columnar phase structure. The strong austenite columns support the incremental load after the ferrite columns yield. For the equiaxed hexagonal grain, random phase microstructure, it is easier for yielded regions to interconnect and form a yield band structure. Microstructure also has a greater effect on fiber-averaged lattice strains when the phases have dissimilar strength, as shown in Figure~\ref{fig:LSMicrostructStrength}. The simulated and experimental data do not match in these plots, because the material parameters have been altered. The experimental data are shown solely for reference. The effects of microstructure on the fiber-averaged lattice strain response become apparent late in the elasto-plastic transition, around 400 MPa, and and persist into the fully-developed plastic regime. Microstructure does not affect the elastic response or the initial inflections in the lattice strain curvature that correspond to the initiation of yielding.

\clearpage

\section{Discussion}
\label{sec:discussion}
The mechanical response predicted using  finite element simulations of virtual polycrystals depends, of course, on how the polycrystal is defined.  The complete definition requires a considerable amount of information  to 
fully describe the mechanical properties  and the geometric arrangement of the constituent phases.  Some attributes more strongly influence the mechanical response than others, and thus those that do play a dominant role merit greater attention than attributes that have only a weak affect.  In multiphase, anisotropic materials like duplex steels, separating the critical attributes from  inconsequential attributes is not trivial as the behaviors depend on the attributes in complex ways.   Quantifying the dependencies requires sensitivity studies together with comparisons to relevant data from physical experiments.     

The studies reported here cover two major aspects of the definition of virtual polycrystals for the LDX2101 duplex steel :  the single crystal mechanical properties and the spatial arrangement of the microstructure.  
Both the macroscopic stress-strain response and crystal-scale lattice strain responses guided the investigations of parameter identification for the single crystal properties and of the instantiations of virtual microstructures. 
The findings for the investigations are summarized in the following subsections.
\subsection{ Material parameter study}
The anisotropy of the single crystal elastic moduli plays a crucial role in the behavior.  
The distribution of stress within the elastic domain and the lattice strain transients observed 
during the elastic-plastic transition are strongly influenced to the strength of the elastic anisotropy.
Published values for both the austenitic and ferritic phases performed well in these regards.  
Single crystal elastic constants are listed in Table~\ref{tab:ElasticConstants}.
\begin{table}[h]
	\centering
	\caption{Single crystal elastic constants using the strength of materials convention ($\tau_{44} = c_{44} \gamma_{44}$).}
	\begin{tabular} {c c c c c}		
		phase & $c_{11}$ & $c_{12}$ & $c_{44}$&   source \\
		& (GPa) & (GPa) & (GPa)  & - \\ \hline
		FCC & 205 & 138 & 126& \cite{Ledbetter01a} \\ 
		BCC & 237 & 141 & 116& \cite{Simmons71a}
	\end{tabular}	
	\label{tab:ElasticConstants}
\end{table}

For the plasticity model, there are two components of the model that must be considered: the fixed-state kinetics relation (also called the slip system flow law) and the evolution of state relations.  The critical parameter for the fixed-state kinetics is the rate sensitivity, $m$.  Based on values reported in the literature and our own strain-rate jump tests, values of 0.02 and 0.013 were determined for the austenitic and ferritic phases, respectively.
The evolution equation for the slip system strengths embodies several parameters that must be defined:
$n^\prime$,  $\dot{\gamma}_0$,   $h_0$,  $g_0$,  $g_s$.  A number of variations were considered that allowed for equal  or unequal initial strength, equal or unequal hardening rate, and different saturation strengths.
When evaluated against both macroscopic stress-strain curves and lattice strain data for multiple fibers in both phases, it was concluded that the simulations that treated the two phases  that same did as well or better than other combinations overall.  Thus, in the spirit of keeping the model as simple as merited by the data, identical parameters were designated for the austenite and ferrite phases.
The final choices for the plasticity parameters are presented in Table~\ref{tab:PlasticityParams}. 
Agreement between simulated and experimental lattice strains across a range of biaxial ratios, presented in \cite{pos_daw_twophase}, demonstrates the robustness of this set of parameters in spite of the simplicity.
\begin{table}[h]
	\centering
	\caption{Plasticity parameters.}	
	\begin{tabular} {c c c c c c c}
		phase & $m$ & $n^\prime$ & $\dot{\gamma}_0$ & $h_0$ & $g_0$ & $g_s$ \\
		& & & ($\mathrm{s^{-1}})$ & (MPa) & (MPa) & (MPa) \\ \hline
		FCC & 0.020 & 1 & $10^{-4}$ & 336 & 192 & 458 \\ 
		BCC & 0.013 & 1 & $10^{-4}$ & 336 & 192 & 458
	\end{tabular}
	\label{tab:PlasticityParams}
\end{table}

\subsection{  Virtual polycrystal instantiation study}

Duplex steels exhibit fairly complex microstructures.  The grain shapes of the austenitic and ferritic phases are qualitatively different (ferrite being more irregular than austenite) due to the nature of the phase transformation that produces the austenite upon cooling.  In addition, the average grain sizes differ between austenite and ferrite, with the ferrite grains typically being several times larger than the austenite grains.  Further, the morphology of the phases is influenced by material's processing.   In the material studied here, the phases were highly elongated in the processing direction in comparison to that observed in a transverse plane.
The extent to which these attributes of the microstructure influence the mechanical behavior was
explored here by considering several types of microstructures:  columnar with unequal grain sizes (rolled), columnar with equiaxed hexagonal grains, random phase distribution with equiaxed hexagonal grains, and a Voronoi tessellation of equiaxed grains.  These four types of microstructures range from one that incorporates all the principal attributes of the duplex microstructure to one that includes none of them.   

Several samples  of each type of microstructure were instantiated and used with the finite element formulation to simulate the behavior of the material over a monotonic tensile loading  to 10\% strain.  For all of the microstructures, the measured textures for both the ferrite and austenite were used to assign lattice orientations to grains.  The results were compared to the macroscopic stress-strain response and to lattice strains for several crystallographic fibers in both the ferrite and austenite.  
The simulations showed  little influence from microstructure for material properties that had been identified for LDX2101.  These properties differ for the elastic moduli, but are similar for the strength and hardening.  
Boosting strength of either phase by an order of magnitude does have an important effect,
but such a large difference is not supported by the existing data for LDX2101.

\begin{figure}[h]
	\centering
	\subfigure[Grain]{\includegraphics[width=0.49\linewidth]{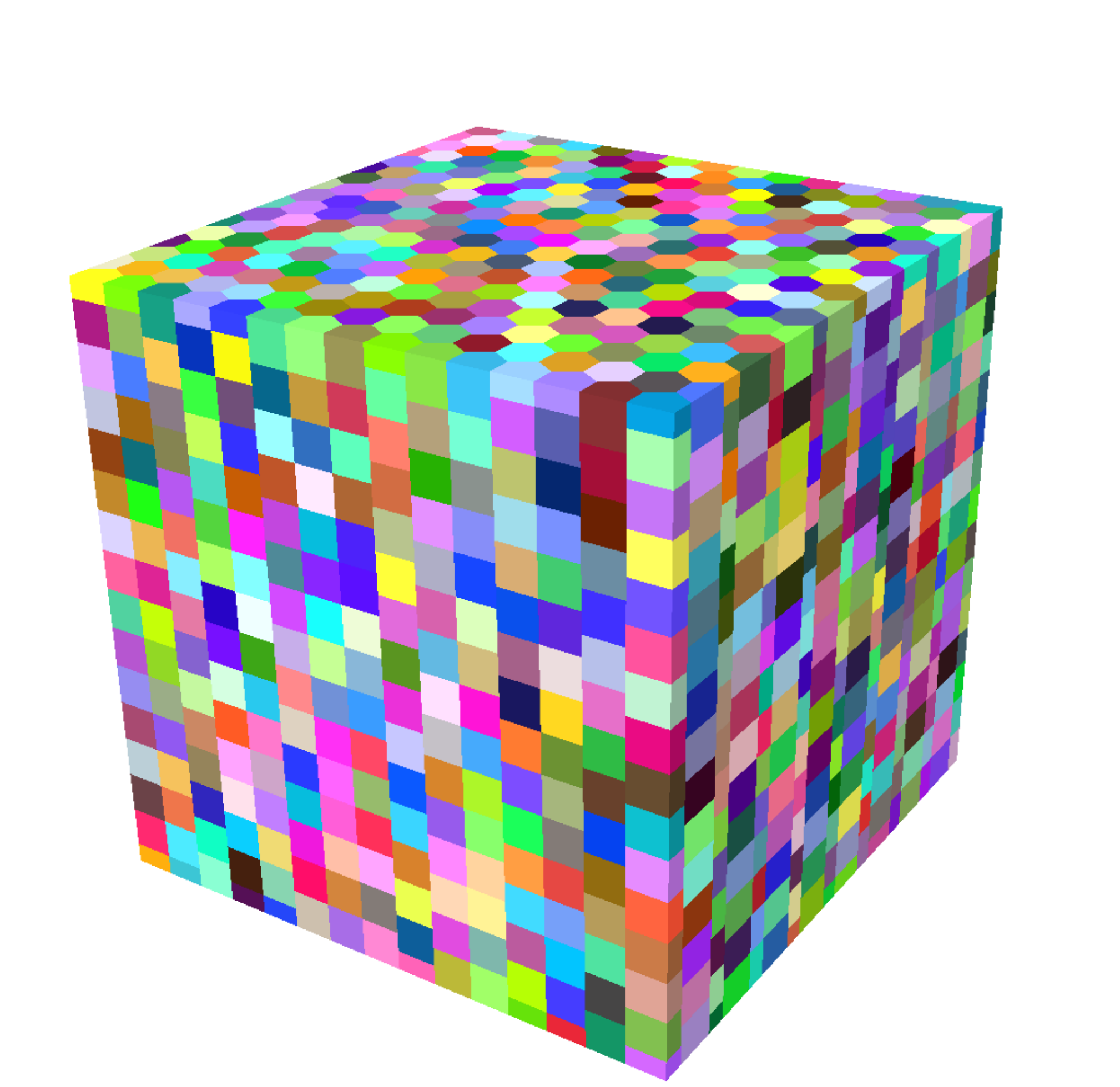}\label{fig:hex15-30-1-1-grain}}
	\subfigure[Phase]{\includegraphics[width=0.49\linewidth]{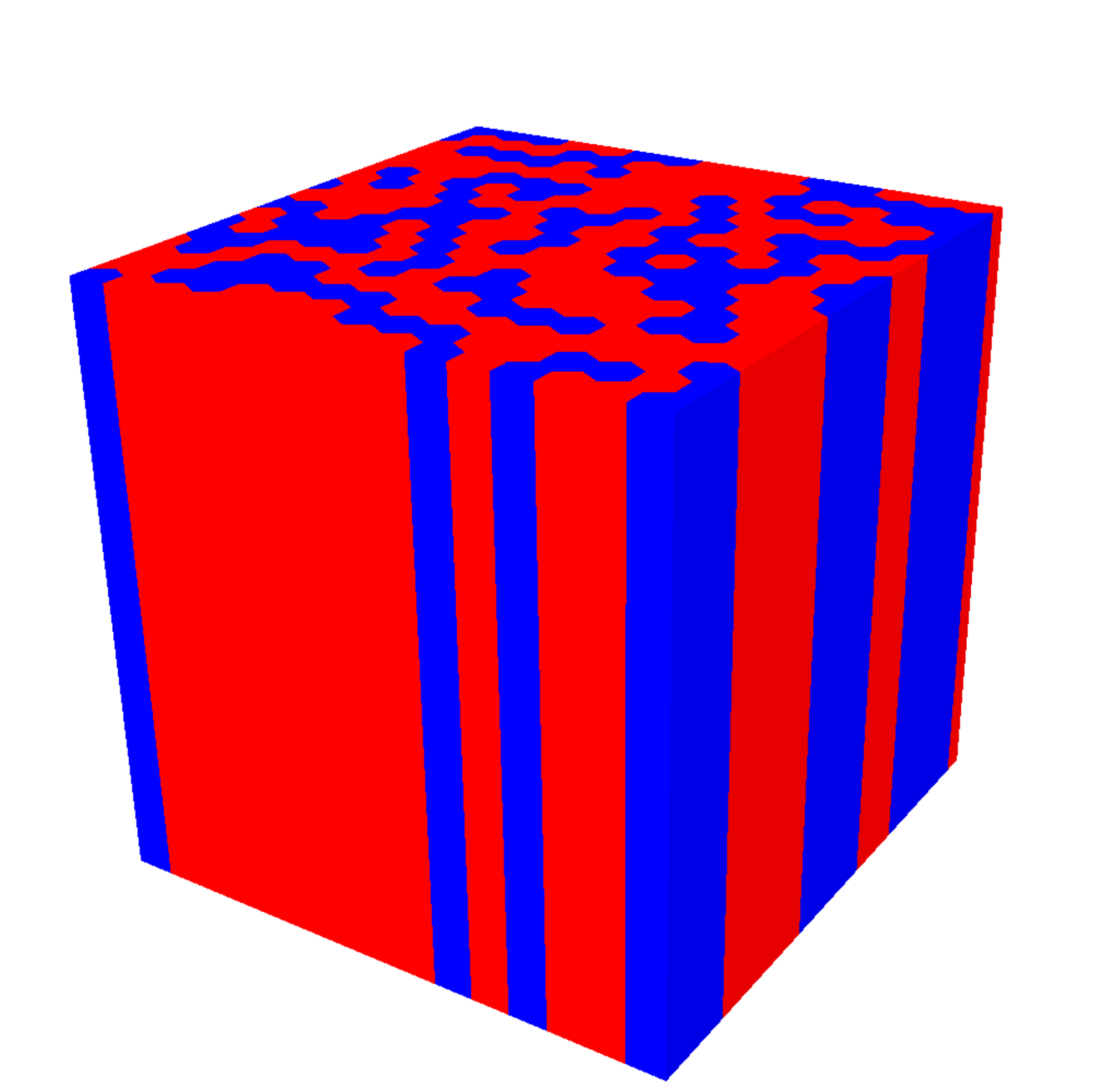}\label{fig:hex15-30-1-1-phase}}
	\caption{Equiaxed virtual microstructure used for the simulation-experiment comparison.}
	\label{fig:SimExpMesh}
\end{figure}

 Modeling the neutron experiments is computationally intensive, due to the complex load history and the large number of grains required to obtain meaningful fiber-averaged statistics.
The results here were used to construct a simplification of the microstructure.  A simplified microstructure was built consisting of equiaxed austenite and ferrite grains, but retaining major features of the phase structure as described above.  This microstructure was used for the simulation-experiment comparison in \cite{pos_daw_twophase}.   The mesh discretization of the microstructure consisted of 137,700 elements and 193,431 nodes, and was comprised of 2,320 austenite grains and 1,753 ferrite grains (Figure~\ref{fig:SimExpMesh}).

\clearpage

\section{Conclusions}
\label{sec:conclusions}

 In this paper we have investigated  sources of uncertainty in defining virtual samples 
 used in simulations of the mechanical behavior of the duplex steel, LDX2101. 
Uncertainty enters in the constitutive parameters and in the definition of virtual samples.
For the constitutive parameters, the simulations require the elastic moduli and parameters
associated with the kinetics and state evolution of the slip systems.
For the virtual samples, the spatial arrangement of the phases (topology and morphology) are needed,
as well as geometric information of the grains, including grain size and shape and lattice orientations.

 For the constitutive parameters, available data included macroscopic stress-strain records for quasi-static tension tests, lattice strain data from {\it in situ} loading, neutron diffraction, and strain-rate increment/decrement tests in addition to properties reported in the literature.  Sets of simulations were performed which mimicked the experimental loading and iterated on constitutive parameter to identify parameter values that provided satisfactory comparisons with observed responses.  Published values for austenite and ferrite gave good comparisons to both the aggregate behavior in the macroscopic stress-strain behavior and the  crystal behavior in the lattice strains within the elastic domain.   It was found that the strain rate sensitivity of the ferritic phase is lower than the austenite phase (0.013 compared to 0.020).  Using the same evolution equation parameters  for both the ferrite and austenite phases provided comparisons between simulated and measured lattice strains that were as good overall as specifying unequal hardening.  This was true for either higher or lower rate in the austenite compared to the ferrite.  
 
 For the instantiation of virtual polycrystals (samples), several types of microstructures were considered: 
columnar phase with unequally-sized grains;
columnar phase with equiaxed grains;
random phase with equiaxed grains; and,
Voronoi tessellation.  
These represented a range of microstructures in terms of their similarities to the actual rolled microstructure with
the columnar phase with unequally-sized grains being most similar and the Voronoi tessellation being the least.
With similar phase strengths and hardening behaviors between the phases (the result of the constitutive parameter identification study), the sensitivity of the computed lattice strains to the type of microstructure was found to be weak.  In contrast,
there was strong sensitivity to the inclusion of anisotropic elastic moduli or to large phase strength contrasts. 
The results were used to construct appropriate virtual samples for a study on the role of strength-to-stiffness on the initiation and propagation of yielding in the duplex steel~\cite{pos_daw_twophase}.

\section{Acknowledgements}
\label{sec:acknowledgements}
Support was provided  by the US Office of Naval Research (ONR) under contract N00014-09-1-0447.
Neutron diffraction experiments were performed on a National Research Council (Canada) neu- tron diffractometer located at the NRU Reactor of AECL (Atomic Energy of Canada Limited).

\bibliographystyle{unsrt}
\bibliography{References}

\end{document}